\documentclass[useAMS,usenatbib,twocolumn]{mn2e}

 \usepackage{epsfig}

\usepackage{amssymb}
\usepackage{graphics}
\usepackage{amsmath}

\title[High Lundquist number magnetic
  reconnection]{Numerical simulations of high Lundquist
  number relativistic magnetic reconnection}

\author[O. Zanotti, M. Dumbser] 
{O. Zanotti$^{1}$\thanks{E-mail:
zanotti@aei.mpg.de}, M. Dumbser$^{2}$ \\
$^{1}$Max-Planck-Institut f{\"u}r Gravitationsphysik, Albert Einstein Institut, Golm, Germany \\
$^{2}$Laboratory of Applied Mathematics, University of
  Trento, Via Mesiano 77, I-38100 Trento, Italy}
\begin{document}

\date{}

\pagerange{\pageref{firstpage}--\pageref{lastpage}} \pubyear{2009}

\maketitle

\newcommand{\oz}[1]{\textcolor{red}    {\texttt{\textbf{OZ: #1}}} }
\newcommand{\be}{\begin{equation}}
\newcommand{\ee}{\end{equation}}
\newcommand{\bdm}{\begin{displaymath}}
\newcommand{\edm}{\end{displaymath}}
\newcommand{\bea}{\begin{eqnarray}}
\newcommand{\eea}{\end{eqnarray}}
\newcommand{\PNM}{P_NP_M}
\newcommand{\halb}{\frac{1}{2}}
\newcommand{\FQi}{\tens{\mathbf{F}}\left(\Qi\right)}
\newcommand{\FQj}{\tens{\mathbf{F}}\left(\Qj\right)}
\newcommand{\FQjj}{\tens{\mathbf{F}}\left(\Qjj\right)}
\newcommand{\nj}{\vec n_j}
\newcommand{\FORCE}{\textnormal{FORCE}}
\newcommand{\GFORCE}{\textnormal{GFORCEN}}
\newcommand{\LF}{\textnormal{LF}'}
\newcommand{\LW}{\textnormal{LW}'}
\newcommand{\WL}{\mathcal{W}_h^-}
\newcommand{\WR}{\mathcal{W}_h^+}
\newcommand{\nur}{\boldsymbol{\nu}^\textbf{r} }
\newcommand{\nuf}{\boldsymbol{\nu}^{\boldsymbol{\phi}} }
\newcommand{\nut}{\boldsymbol{\nu}^{\boldsymbol{\theta}} }
\newcommand{\ar}{\phi_1\rho_1}
\newcommand{\arr}{\phi_2\rho_2}
\newcommand{\ur}{u_1^r}
\newcommand{\uf}{u_1^{\phi}}
\newcommand{\ut}{u_1^{\theta}}
\newcommand{\urr}{u_2^r}
\newcommand{\uff}{u_2^{\phi}}
\newcommand{\utt}{u_2^{\theta}}
\newcommand{\ub}{\textbf{u}_\textbf{1}}
\newcommand{\ubb}{\textbf{u}_\textbf{2}}
\newcommand{\RoeMat}{{\tilde A}_{\Path}^G}

\label{firstpage}

\begin{abstract}
We present the results of 
two-dimensional and three-dimensional 
magnetohydrodynamical numerical simulations of
relativistic magnetic reconnection,
with particular emphasis on the dynamics of the plasma
in a Petschek-type configuration with high Lundquist
numbers, $S\sim 10^5-10^8$.
The numerical scheme adopted, allowing for 
unprecedented accuracy for this type of calculations, 
is based on high order finite volume 
and discontinuous Galerkin methods as recently proposed by
\citet{Dumbser2009}.  
The possibility of producing high Lorentz factors is
discussed, showing
that Lorentz factors close to $\sim 4$ can be
produced for a  plasma parameter $\sigma_m=20$.
Moreover, we find that the Sweet-Parker layers are
unstable, generating secondary magnetic islands,
but only for $S>S_c\sim 10^8$, much larger than 
what is reported in the Newtonian regime.
Finally, the effects of a mildly anisotropic Ohm law are 
considered in a configuration with a guide magnetic
field. Such effects produce only slightly faster reconnection
rates and Lorentz factors of about $1\%$ larger with respect to
the perfectly isotropic Ohm law.
\end{abstract}


\section{INTRODUCTION}
\label{Introduction}

Relativistic magnetic reconnection is recognized to be a key
physical process in high-energy astrophysics, being able
to convert magnetic energy into heat and plasma kinetic energy
over short timescales.
Relevant examples include:
(i) the magnetospheres of pulsars near the
Y-point, where the outermost magnetic field lines
intersect the equatorial plane \citep{Uzdensky2003,Gruzinov2005}; 
(ii)  the dissipation of alternating fields 
at the termination shock of a relativistic striped pulsar
wind \citep{Petri2007}; 
(ii) soft gamma-ray repeaters, where giant magnetic 
flares could be the explanation 
of the observed strongly magnetized and relativistic
ejection events \citep{Lyutikov2003,Lyutikov2006}; 
(iv) gamma-ray burst jets, where particle acceleration by
magnetic reconnection in electron-positron
plasmas is supposed to take place
\citep{Drenkhahn2002,Barkov2010,McKinney2010,Rezzolla:2011}; 
(v) and accretion disc coronae of 
active galactic nuclei, where 
violent releases of energy may be generated by
Petschek magnetic reconnection of strong magnetic loops emerging
from the disc via buoyancy instability\footnote{
A similar mechanism has been
proposed by \citet{Tanuma2003} to explain
the X-ray emission in the Galactic
halo.}~\citep{diMatteo1998,Schopper1998,Jaroschek2004}. 
Very recently, moreover, \cite{Nalewajko2011} have described
a model of mini jets in blazars to explain their ultra-fast
variability and were able to fit data of PKS2155-304
assuming that the dynamics is governed by relativistic
magnetic reconnection with a weak 
``guide magnetic field'', i.e. with a magnetic field
aligned to the current. 

In the recent past, both theoretical studies and
numerical investigations have greatly improved our
understanding of relativistic magnetic reconnection. 
Although within the incompressibility assumption,
\citet{Lyutikov2003b} first found that 
in the Sweet-Parker reconnection three different
regimes can be produced, which depend on
the ratio of the magnetization parameter $\sigma_m$ 
to the Lundquist number $S$.
\cite{Lyubarsky2005}, on the other hand, explicitly
addressed the question about 
whether the reconnection rate can be significantly
enhanced or not in the transition from
Newtonian to relativistic magnetic reconnection.
He found that, unless the reconnecting fields are not
strictly antiparallel, 
the relativistic Petschek reconnection should
not be considered as a mechanism for the direct
conversion of the magnetic energy into the plasma energy
and the reconnection rate would be at most $0.1$ the
speed of light, contrary to what originally suggested by
\citet{Blackman1994}.
Moreover, \citet{Tolstykh2007} extended the Petschek reconnection
model to incorporate relativistic effects of impulsive
reconnection and, for current layers embedded into
strong magnetic fields, 
they claimed that the plasma can be  accelerated to
high Lorentz factors.

Because our understanding of relativistic magnetic
reconnection is still incomplete, in the last few years
there has been a growing expectation towards numerical
simulations as a promising tool for clarifying the rich
underlying physics, particularly in the non-linear regime.
After the pioneering resistive relativistic simulations
by \cite{Watanabe2006}, who considered the Petschek type
reconnection in the relativistic regime with a
resistive magnetohydrodynamic (MHD) code,
\citet{Zenitani2009} investigated the relativistic
reconnection in an electron-positron plasma by  a
two-fluid MHD code
and showed that the reconnection rate is
higher and higher in magnetically dominated regimes.
\citet{Zenitani2009_guide}, on the other hand, clarified
the role of a guide magnetic field, 
which is essentially that of making the output
energy flux Poynting dominated rather than enthalpy
dominated. Very recently, \cite{Zenitani2010} showed
that, when the resistivity is current dependent,
plasmoids are repeatedly formed in the current sheet.
In spite of these progresses, two major numerical limitations
still prevent the application of numerical
calculations to realistic physical and astrophysical
scenarios in the relativistic regime. 
The first limitation manifests when treating plasmas with 
high magnetizations parameters $\sigma_m$, while the
second limitation manifests when high Lundquist numbers
$S$ are encountered, being related to the stiffness of
the equations in this regime. Physically, 
high Lundquist number plasmas are very interesting as
they are supposed to become unstable and break up into
secondary magnetic islands.
In the Newtonian framework, for example,
\citet{Samtaney2009} showed that Sweet-Parker current
sheets are unstable to super-Alfvenically fast formation
of plasmoid chains for $S>S_c\sim 10^4$, and  it is not
clear the extent to which this instability persists in
a relativistic framework.

In this paper, by adopting the innovative numerical method
presented in~\citet{Dumbser2009}, we reconsider some of
the scenarios discussed above within the relativistic
magnetohydrodynamic single
fluid approximation, focusing, in particular, on the
dynamics of the plasma in a regime
characterized by high Lundquist numbers, $S\sim 10^5-10^8$
and mild magnetization parameters $\sigma_m\sim 1-20$.

We also
investigate numerically, for the first time, the effect of an
anisotropic Ohm law, i.e. when the
conductivity of the plasma is a tensor that depends on the orientation of
the magnetic field.

The plan of the paper is the following. In
Section~\ref{sec:Mathematical_formulation} we report the
relativistic resistive equations and the basic physical
assumptions, while Section~\ref{sec:Numerical_method} is
devoted to a succinct presentation of the numerical
method and to the validation of the code in three space
dimensions. Section~\ref{sec:Magnetic_Reconnection}
contains the results of our analysis, and
Section~\ref{sec:Conclusions} the conclusions.
We have considered only flat spacetimes in pseudo-Cartesian
coordinates, namely the metric 
$\eta_{\mu\nu}={\rm diag}(-1,1,1,1)$, where from now onwards 
we agree to use Greek letters
$\mu,\nu,\lambda,\ldots$ (running from 0 to 3) for
indices of four-dimensional space-time tensors, while
using Latin letters $i,j,k,\ldots$ (running from 1 to 3) 
for indices of three-dimensional spatial tensors.
We set the speed of light $c=1$ and make use of the
Lorentz-Heaviside notation for the electromagnetic
quantities, such that all $\sqrt{4\pi}$ factors disappear.  
Finally, we use Einstein summation convention over
repeated indices.

\section{Mathematical formulation}
\label{sec:Mathematical_formulation}

The total energy-momentum tensor of the plasma that we consider
is made up by two contributions,
$T^{\mu\nu}=T^{\mu\nu}_{m}+T^{\mu\nu}_{f}$. The first one
is due to matter 
\be
T^{\mu\nu}_{m}=h\rho\,u^{\,\mu}u^{\nu} + p\eta^{\,\mu\nu},
\label{eq:T_matter}
\ee
where $u^\mu$ is the four velocity of the fluid, while
$h$, $\rho$ and $p$ are  the specific enthalpy, the rest
mass density and the pressure as
measured in the co-moving frame of the fluid. The second
contribution comes from the electromagnetic field
\be
T^{\mu\nu}_{f}={F^{\mu}}_{\lambda}F^{\nu\lambda}-\textstyle{\frac{1}{4}}(F^{\lambda\kappa}F_{\lambda\kappa})\eta^{\,\mu\nu}
\ ,
\label{eq:T_field}
\ee
where $F^{\mu\nu}$, and its dual
$F^{\ast\mu\nu}$, is the electromagnetic tensor given by
\bea
\label{emtensor1}
F^{\mu\nu} &=& n^{\,\mu}E^{\nu} - E^{\mu}n^{\nu} +
\epsilon^{\,\mu\nu\lambda\kappa}B_{\lambda}n_{\kappa} \\
\label{emtensor2}
F^{\ast\mu\nu} &=& n^{\,\mu}B^{\nu} - B^{\mu}n^{\nu} -
\epsilon^{\,\mu\nu\lambda\kappa}E_{\lambda}n_{\kappa} \ .
\eea
$E^{\nu}$ and  $B^{\nu}$ are the electric and
magnetic field as measured by the observer defined by the
four-velocity vector $n^\mu$, while
$\epsilon^{\,\mu\nu\lambda\kappa}=[\mu\nu\lambda\kappa]$
is the completely antisymmetric spacetime Levi-Civita
tensor, with the convention that $\epsilon^{\, 0 1 2 3}=1$.
If we now set $n^\mu$ to define the inertial laboratory observer,
namely  $n^\mu=(1,0,0,0)$, normalized such that $n^\mu
n_\mu=-1$, then the four vectors of the
electric and of the magnetic field are purely spatial, i.e.
$E^0=B^0=0$, $E^i=E_i$, $B^i=B_i$ in this frame.
On the other hand, the fluid four velocity
$u^\mu$ and the standard three velocity in the laboratory
frame are related as $\vec v = v^i=u^i/\Gamma$, where
$\Gamma=({1-\vec v^{\, 2}})^{-1/2}$ is the 
Lorentz factor of the fluid with respect to the
laboratory frame.

In Cartesian coordinates, using the abbreviations 
$\partial_t = \frac{\partial}{\partial t}$ and 
$\partial_i = \frac{\partial}{\partial x_i}$, the full
system  of Euler and Maxwell equations are
\bea
\label{fluid1}
&& \partial_t D + \partial_i (D v^i)=0, \\
\label{fluid2-4}
&&\partial_t S_j + \partial_i Z_{j}^i=0, \\
\label{fluid5}
&&\partial_t \tau + \partial_i S^i=0, \\
\label{electric6-8}
&&\partial_t E^i - \epsilon^{ijk}\partial_j B_k + \partial_i
\Psi = -J^i, \\
&&\partial_t B^i + \epsilon^{ijk}\partial_j E_k + \partial_i
\Phi = 0, \\
\label{divE}
&&\partial_t \Psi + \partial_i E^i = \rho_c - \kappa \Psi,
\\
\label{divB}
&&\partial_t \Phi + \partial_i B^i = - \kappa \Phi, \\
\label{charge}
&&\partial_t \rho_c + \partial_i J^i = 0, 
\eea
where the conservative variables of the fluid are
\bea
D&=&\rho \Gamma, \\
S^i & = & \omega \Gamma^2 v^i + \epsilon^{ijk}E_jB_k, 
\label{eq:S} \\
\tau & = & \omega \Gamma^2 - p +
\textstyle{\frac{1}{2}}(E^2+B^2) \ ,
\label{eq:U}
\eea
expressing, respectively, the relativistic mass density, the
momentum density and the total energy density.
The spatial tensor $Z^i_j$ in (\ref{fluid2-4}), representing the momentum 
flux density, is 
\bea
\label{eq:W} 
Z^i_j & = & \omega \Gamma^2 v^i\,v_j -E^i\,E_j-B^i\,B_j+ \nonumber
\\
&&\left[p+\textstyle{\frac{1}{2}}(E^2+B^2)\right]\,\delta^i_j,
\eea
where $\omega=h\rho$ is the enthalpy of the fluid while
$\delta^i_j$ is the Kronecker delta. 
An equation of state is needed to close the system, 
and we have chosen that of an ideal gas, namely
\be
\label{eos}
p=(\gamma-1) \rho\epsilon=\gamma_1(\omega-\rho), 
\ee
where $\gamma$ is the adiabatic index,
$\gamma_1=(\gamma-1)/\gamma$, $\epsilon$ is the specific internal energy.

While writing equations \eqref{divE} and
\eqref{divB}, we have adopted the so called 
{\em divergence-cleaning
approach} presented in \cite{Dedner2002}, which amounts to
introducing two additional scalar fields $\Psi$ and
$\Phi$ that propagate away the deviations of the
divergences of the electric and of the magnetic fields
from the values prescribed by Maxwell's equations.
Additional details about this approach 
can be found in \cite{Komissarov2007} and
\cite{Palenzuela:2008sf}.

In its most general form, the relativistic
formulation of Ohm law is a non-linear propagation equation
\citep{Kandus2008}, which makes it manifest the connection
between the Ohm law and the equations of motion. 
However, in
order to keep the equations numerically tractable and
because of the poor knowledge of the conductivity in
realistic conditions,
simpler forms of the Ohm law are usually considered, in
which the currents are algebraically related to the
electromagnetic field. Following \cite{Bekenstein1978}
we allow for an anisotropic Ohm law and therefore we write
the four-current vector as
\be
\label{ohm1}
I^\mu=q_0 u^\mu + \sigma^{\mu\nu}e_\nu, 
\ee
where $q_0$ is the charge density in the co-moving
frame of the plasma, $\sigma^{\mu\nu}$ is the conductivity tensor and
$e_\nu$ is the electric field in the co-moving frame. 
Within the collision-time approximation the most general form
of the conductivity tensor is given by
\be
\label{ohm2}
\sigma^{\mu\nu}=\sigma(g^{\mu\nu}+\xi^2
b^{\mu}b^{\nu}+\xi \epsilon^{\mu\nu\lambda\kappa}u_\lambda
b_\kappa)
\ee
where $u^\mu$ is the four velocity of the plasma
and $b^\mu$ is the magnetic field in the
co-moving frame. The parameter $\xi$
is related to the micro-physics of the
plasma~\citep{Bekenstein1978} via $\xi=e\tau/m_e$, where 
$e$ and $m_e$ are electron's charge and mass, while
$\tau$ is the collision time.
As a first application to the case of an anisotropic Ohm
law, in this paper we consider the case
$\sigma^{\mu\nu}=\sigma(g^{\mu\nu}+\xi^2
b^{\mu}b^{\nu})$, namely we drop the third term
in (\ref{ohm2}). 
We note, therefore, that anisotropic effects in the Ohm law are
more and more important as the intensity of the magnetic
field is increased. 
The four-current vector can also be decomposed in
components parallel and perpendicular to the 
observer as $I^\mu=\rho_c n^\mu + J^\mu$, where
$\rho_c$ is the charge density in the laboratory
frame. Hence, it is easy to derive from (\ref{ohm1}) (see
Appendix~\ref{appendixA} for the calculations) the following
expression for the spatial current vector to be used
in~Eq.(\ref{electric6-8}), namely 
\bea
\label{ohmlaw}
\vec J &=& \rho_c \vec v + \Gamma \sigma [ \vec E + \vec v
\times \vec B - (\vec E \cdot \vec v)\vec v] + \nonumber \\
&&
\Gamma \sigma \xi^2 (\vec E \cdot \vec B)
[\vec B - \vec v \times \vec E - (\vec B \cdot \vec
  v)\vec v]  \ .
\eea
Two relevant comments are worth doing about the Ohm law
(\ref{ohmlaw}). The first and most obvious one is that 
the isotropic regime is recovered when 
$\xi=0$, in which case Eq.~(\ref{ohmlaw})
reduces to the usual expression~\citep{Komissarov2007}. 
The second comment is that, in the
comoving frame, the Ohm law (\ref{ohmlaw}) becomes
\be
\label{ohmlaw2}
\vec J=\sigma \vec E + \sigma\xi^2(\vec E \cdot \vec
B)\vec B
\ee
which clarifies how the current term proportional to $\xi^2$ 
is present only for
configurations for which $\vec E \cdot \vec B\neq0$ and
it is responsible for an extra current term in the direction
parallel to the magnetic field.

\section{Numerical method}
\label{sec:Numerical_method}

%
\subsection{Brief description}

A well known and challenging feature of 
the system of equations (\ref{fluid1})-(\ref{charge}) is
that it has
source terms in the three equations (\ref{electric6-8})
for the evolution of the electric field
that become stiff  in the limit of high conductivity.
This pathology has been handled in the recent past by
resorting to Strang-splitting techniques
\citep{Komissarov2007} or to implicit-explicit Runge
Kutta methods~\citep{Palenzuela:2008sf}. In the present
paper, on the other hand, we apply the strategy outlined
by \citet{Dumbser2009}, who used the so called high order 
${P_NP_M}$ methods, which are a unification of high order 
finite volume and discontinuous Galerkin finite element
schemes in a more general framework, and combining 
results from \cite{DET2008} and \cite{DBTM2008}.
It is worth stressing that, because of the high accuracy they allow to achieve,
Galerkin methods have been recently considered as a
valuable approach even in fully relativistic calculations. 
Promising results have been obtained by \citet{zumbusch_2009_fed}
and \cite{Radice2011}.

\begin{figure}
\psfig{file=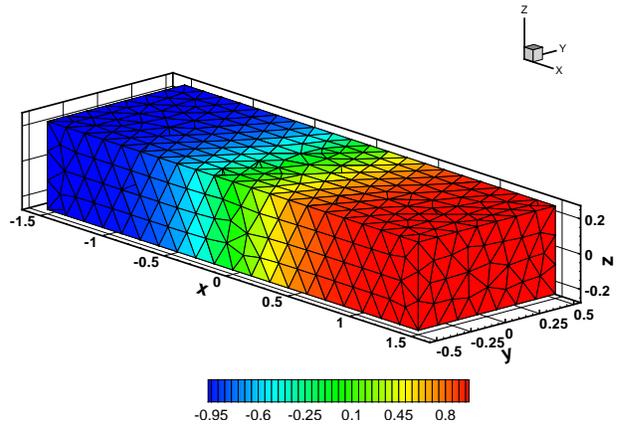,width=8.5cm}
\caption{Numerical grid of the 3D simulation in the
  current sheet test. The $B^y$ component of the magnetic
  field is reported.}
\label{Komissarov3D}
\end{figure}
\begin{figure}
\psfig{file=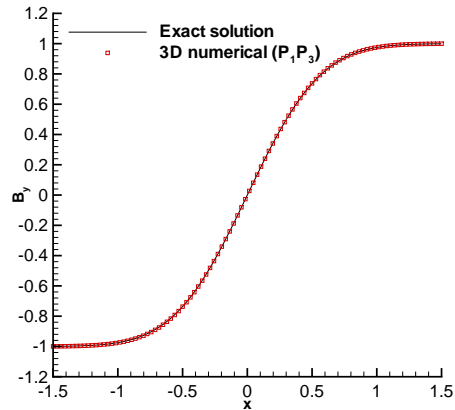,width=6.5cm}
\caption{Comparison of the numerical solution with the
  analytic one in the 3D current sheet test. The plot
  shows a one-dimensional cut of the magnetic field $B^y$
at time $t=10$.}
\label{Komissarov3D_comparison}
\end{figure}
%

\begin{table*}
\begin{center}
\caption{Main properties of the initial models. From left to right the
  columns report the name of the model, the magnetization
  parameter $\sigma_m$, 
  the Lundquist number $S$, the background resistivity $\eta_b$,
  the value of the vertical magnetic field $B_z$,
  the number of cells used and the minimum cell size
  $L_{\rm min}$
  (used for time-step calculation).
  All of the
  models have adiabatic index $\gamma=4/3$, $p_0=1$,
  $\rho_0=1$ and $\eta_{i0}=1.0$. A  
  third order $P_0P_2$ finite volume scheme has been adopted.
}
\label{tab1}
\begin{tabular}{l|cccccc}
\hline
\hline
Model & $\sigma_m$  & $S$ & $\eta_b$  & $B_z/B_0$ &cells & $L_{\rm min}$\\

\hline

\texttt{2D-m1-S1.60e5} & $1.0$ & $1.60\times10^5$  &$10^{-3}$ & $0.0$ &$844194$ & $4.90\times10^{-3}$\\
\texttt{2D-m5-S2.45e5} & $5.0$ & $2.45\times10^5$  &$10^{-3}$ & $0.0$ &$844194$ & $4.90\times10^{-3}$\\
\texttt{2D-m10-S2.68e5} & $10.0$ & $2.68\times10^5$ &$10^{-3}$& $0.0$ &$844194$ & $4.90\times10^{-3}$\\
\texttt{2D-m15-S2.77e5} & $15.0$ & $2.77\times10^5$ &$10^{-3}$& $0.0$ &$844194$ & $4.90\times10^{-3}$\\
\texttt{2D-m20-S2.82e5} & $20.0$ & $2.82\times10^5$ &$10^{-3}$& $0.0$ &$844194$ & $4.90\times10^{-3}$\\

\hline

\texttt{2D-m10-S2.68e6} & $10.0$ & $2.68\times10^6$ &$10^{-4}$& $0.0$ &$844194$ & $4.90\times10^{-3}$\\
\texttt{2D-m10-S2.68e7} & $10.0$ & $2.68\times10^7$ &$10^{-5}$& $0.0$ &$844194$ & $4.90\times10^{-3}$\\
\texttt{2D-m10-S2.68e8} & $10.0$ & $2.68\times10^8$ &$10^{-6}$& $0.0$ &$844194$ & $4.90\times10^{-3}$\\

\hline

\texttt{3D-m1.25-Bz0} & $1.25$ & $1.73\times10^4$  &$0.005$  & $0.0$ &$6105345$ & $2.49\times10^{-2}$\\
\texttt{3D-m1.25-Bz0.5} & $1.25$ & $1.73\times10^4$ &$0.005$& $0.5$ &$6105345$ & $2.49\times10^{-2}$\\

\hline
\hline

\end{tabular}
\begin{flushleft}

\end{flushleft}
\end{center}
\end{table*}
In our specific implementation, the numerical solution of the vector
of conserved quantities is represented at the beginning of
each time-step by polynomials of degree $N$. However, the
time evolution of these data and the computation of the corresponding
numerical fluxes are performed with a different set of
piecewise polynomials of degree $M\geq N$, which are
reconstructed starting from the underlying $N$ degree
polynomials.  
The part of the algorithm performing the time evolution
of the reconstructed polynomials uses a local
space-time discontinuous Galerkin (DG) finite element
scheme which provides 
a \textit{local predictor} for
constructing the solution of the partial differential
equations \textit{in the
small}, as it was called by \cite{eno}. Our local space--time
DG predictor allows us to discretize the stiff source 
terms arising in the resistive relativistic MHD equations naturally 
and without using any splitting techniques.  
The local space-time predictors are obtained after
solving, for each individual cell element,
a system of non-linear equations, which are therefore
not globally coupled as in the global and
fully implicit space-time Galerkin approach introduced by
\citet{spacetimedg1} and by~\citet{KlaijVanDerVegt}. 
Once computed, the local space-time predictors
 are then inserted into a 
\textit{global corrector}, which is fully explicit  
and provides the coupling between neighboring cells. 
The resulting Galerkin scheme can allow for an 
arbitrary (at least in principle) order of
convergence, through a one-step and quadrature free time update 
(no need for Runge Kutta time stepping) formula.
Within this new approach, traditional finite volume schemes
with $N=0$ and usual discontinuous Galerkin methods with
$N=M$ are included as special cases, while the new class
of methods with $N\neq 0, M>N$ are in general
computationally more efficient. In most of the numerical 
simulations reported in this paper we have
adopted the schemes ${P_0P_2}$, ${P_1P_2}$ and ${P_1P_3}$.

\begin{figure*}
\centering
{\includegraphics[angle=0,width=7.5cm,height=10.0cm]{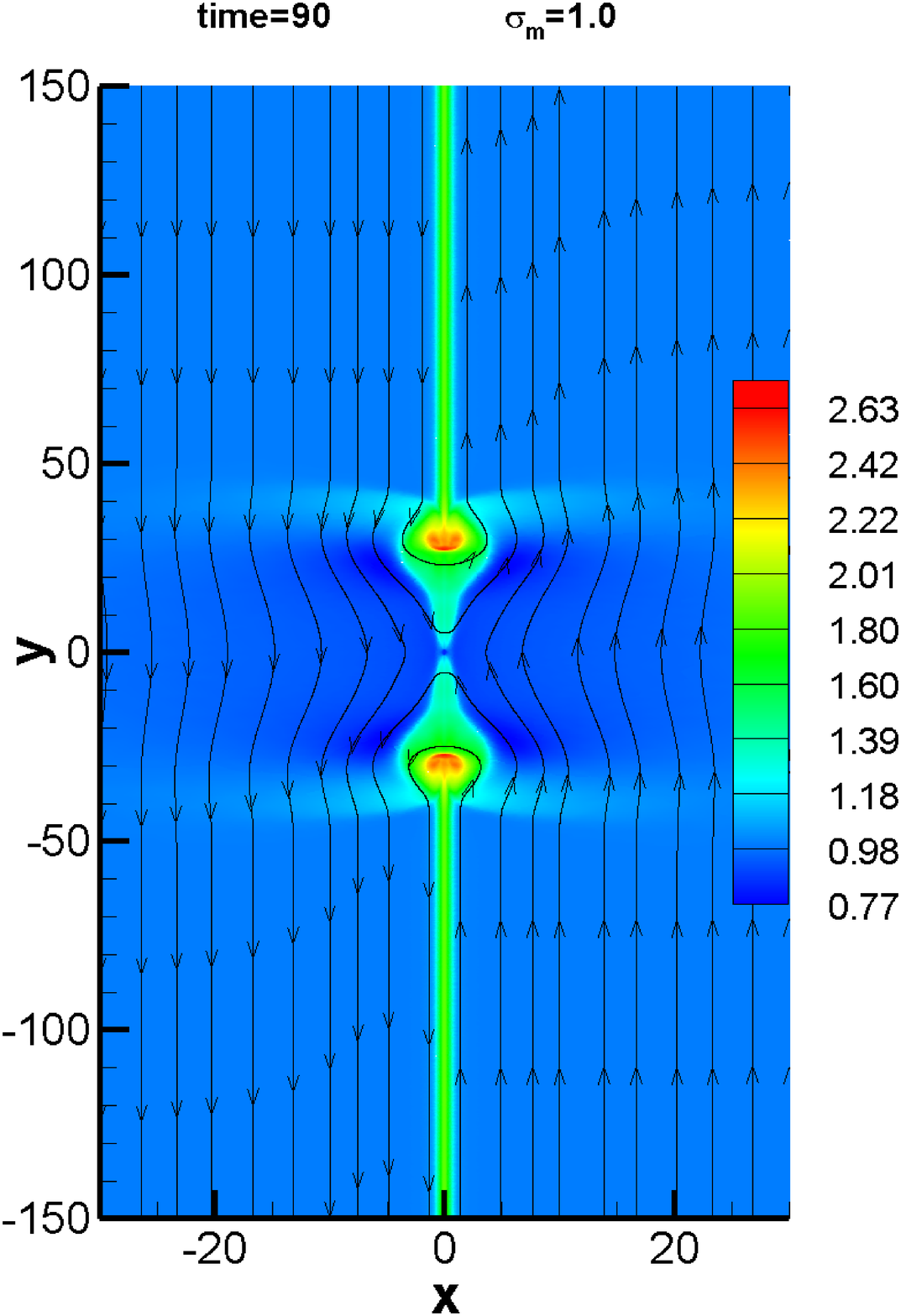}}
\hspace{1cm}
{\includegraphics[angle=0,width=7.5cm,height=10.0cm]{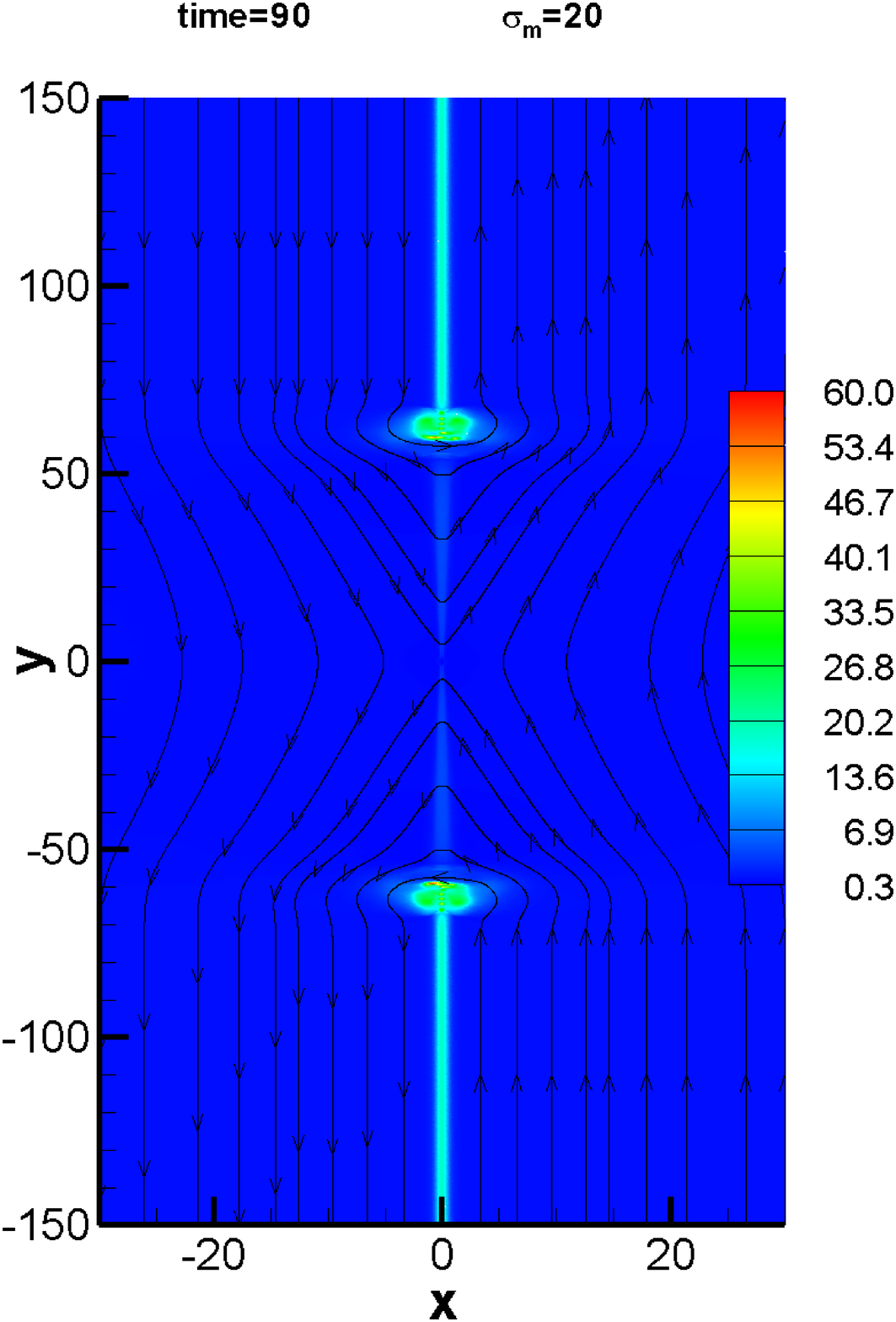}}
\caption{Color map of the rest mass density for two models
  with the same background electrical resistivity $\eta_b=10^{-3}$
  but different
  values of the magnetization: $\sigma_m=1.0$ (left panel) 
  and   $\sigma_m=20$ (right panel). Note that the aspect
  ratio of the figure is not one.}
\label{rho_plus_field_map}
\end{figure*}
%

\subsection{Validation of the code in three dimensions}

In \citet{Dumbser2009} we presented a wide class of
numerical tests both in one and in two spatial dimensions,
showing the ability of the scheme in dealing with the
stiff terms inherent in the resistive relativistic
equations, while retaining its high order properties. 
Here we complete the full validation of the code by
considering the numerical evolution of a self similar
current sheet 
in three spatial dimensions.
This configuration, first proposed by \cite{Komissarov2007}, has
the following exact analytical solution for the $y$-component 
of the magnetic field:
\begin{equation}
\label{eqn.komissarov.exact}
  B_y(x,t) = B_0 \ \text{erf} \left( \frac{1}{2}
  \sqrt{\frac{\sigma}{t}} x \right) \ ,
\end{equation} 
where ${\rm erf}$ is the error function.
The initial time for this test case is chosen to be $t=1$ and the
initial condition at $t=1$ is given by  $\rho=1$, $p=50$,
$\vec E = \vec v = 0$ and $\vec B = (0,B_y(x,1),0)^T$. 
We choose $\gamma = \frac{4}{3}$ and $B_0 = 1$. 
The conductivity is set to $\sigma = 100$, which means a
moderate resistivity. We have solved the problem with
the fourth order $P_1P_3$ scheme on a very coarse 
unstructured mesh composed by 3209 tetrahedrons. 
The grid extension is given by $[-1.5,1.5]\times[-0.5,0.5]\times[-0.25,0.25]$
and it is shown in Fig.~\ref{Komissarov3D}
with a color rendering of the component $B_y$ of the
magnetic field. 
Fig.~\ref{Komissarov3D_comparison}, on the other hand, shows the perfect
matching of the numerical solution against the analytic
one at time $t=10$ and computed along a representative
one-dimensional cut of the numerical domain. We have also
verified that the method provides the expected
order of accuracy when the order of the
polynomials in the $P_NP_M$ scheme is changed. This concludes the
validation of the code in the full three-dimensional case.

\section{Magnetic Reconnection} 
\label{sec:Magnetic_Reconnection} 

\subsection{Initial model and boundary conditions}

The initial model that we have considered in our analysis
of relativistic magnetic reconnection is built on Harris
model, as reported by  \cite{Kirk2003},
and it reproduces a current sheet
configuration in the $x-y$ plane.
Very similar configurations have been studied also by
\citet{Watanabe2006} and by \citet{Zenitani2009_guide,Zenitani2009}.
Gas pressure and density are given by
$p=p_0+\sigma_m\rho_0[p_0\cosh^2(2x)]^{-1}$, 
$\rho=\rho_0+\sigma_m\rho_0[p_0\cosh^2(2x)]^{-1}$, where $p_0$ and
$\rho_0$ are the constant values outside the current
sheet, whose thickness is $\delta=1$.
The magnetic field changes orientation across the
current sheet according to $B_y=B_0\tanh(2x)$, where 
the value of $B_0$ is given in terms of the magnetization parameter
$\sigma_m=B_0^2/(2\rho_0\Gamma_0^2)$.
All over the grid there is a small background uniform resistivity
$\eta_b$,
except for a circle of radius
$r_{\eta}=0.8$, defining a region of anomalous
resistivity of amplitude $\eta_{i0}=1.0$.
As a result, the resistivity can be written as
\begin{eqnarray}
\eta=\left\{
\begin{array}{lr}
{\eta_b +\eta_{i0}\left[2(r/r_{\eta})^3-3(r/r_{\eta})^2+1\right]} & {\rm for} \hspace{12pt} r\leq r_{\eta}, \\
\eta_b & {\rm for} \hspace{12pt} r>r_{\eta}, \\
\end{array}
\right.
\end{eqnarray}
where  $r=\sqrt{x^2+y^2}$.
The velocity field is
initially zero, hence $\Gamma_0=1$, 
while the electric field is given by
$E_z=\eta(\partial B_y/\partial x)$.  
In most of our simulations we have considered the case
with $p_0=1$, $\rho_0=1$.
The numerical grid consists of  an
unstructured mesh composed of triangles in 2D and
tetrahedrons in 3D, which are 
clustered along the current sheet.
The grid extension is given by
$[-50,50]\times[-150,150]$ in 2D and  
by $[-50,50]\times[-75,75]\times[-12.5,12.5]$ in 3D.
Tab.~\ref{tab1} reports the basic parameters
of the models studied in our simulations. 
The Lundquist
number $S=v_AL/\eta_b$ is reported in the  fourth
column, where $L$ is the length of the initial current
sheet, while $v_A^2=B^2/(h\rho+B^2)$ is the relativistic
Alfven velocity~\citep{Komissarov1997,DelZanna2007}.  
The name of the models have been chosen to facilitate 
recognizing their main parameters. For example, model 
$\texttt{2D-m10-S2.68e5}$
is a two dimensional model
with magnetization $\sigma_m=10$ and Lundquist number $S=2.68\times10^5$
(corresponding to $\eta_b=10^{-5}$). 
We have used
periodic boundary conditions at $y_{\rm min}$ and $y_{\rm max}$, while
zeroth order extrapolation is applied at $x_{\rm min}$
and $x_{\rm max}$.

\subsection{Results}

%
\begin{figure}
\psfig{file=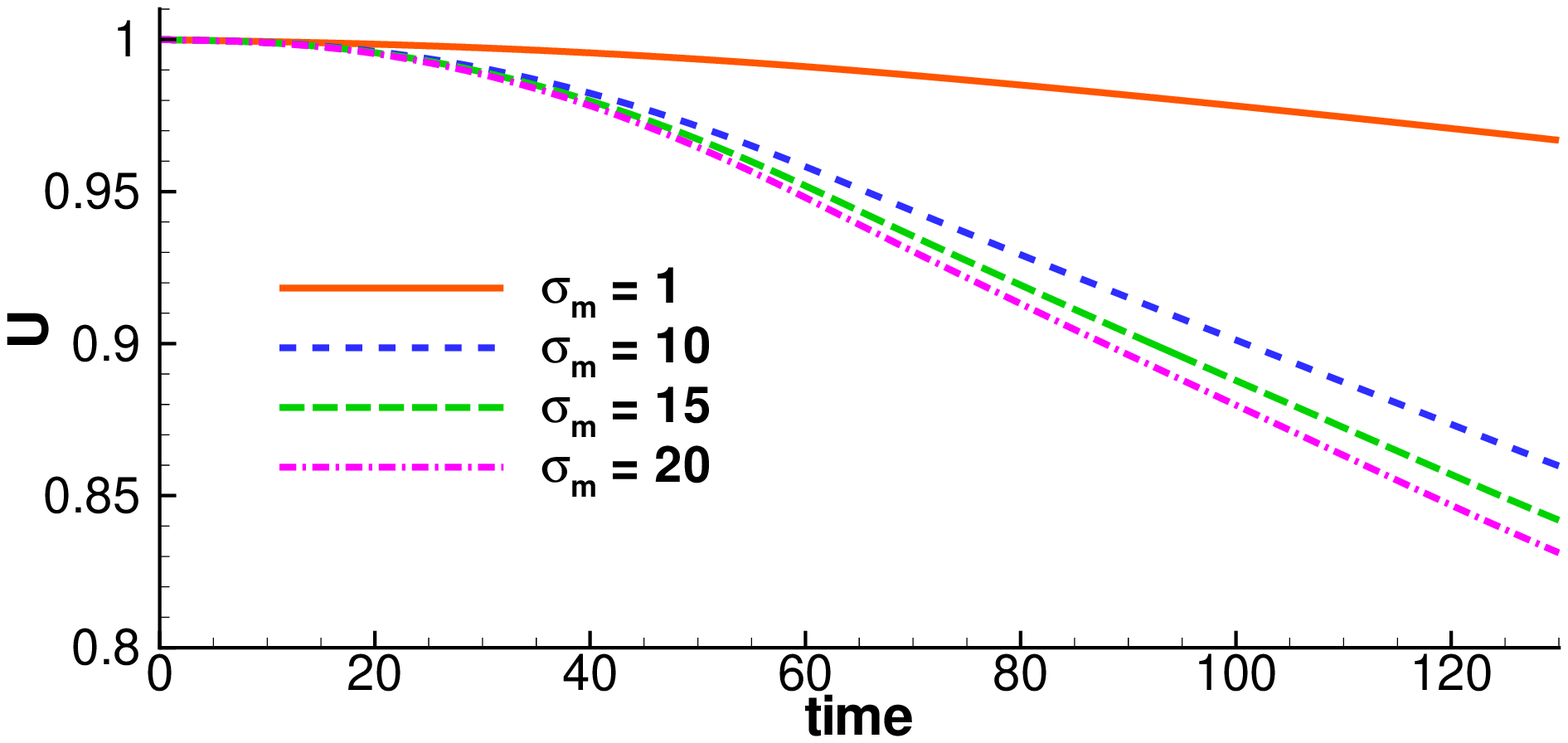,width=8.0cm}
\psfig{file=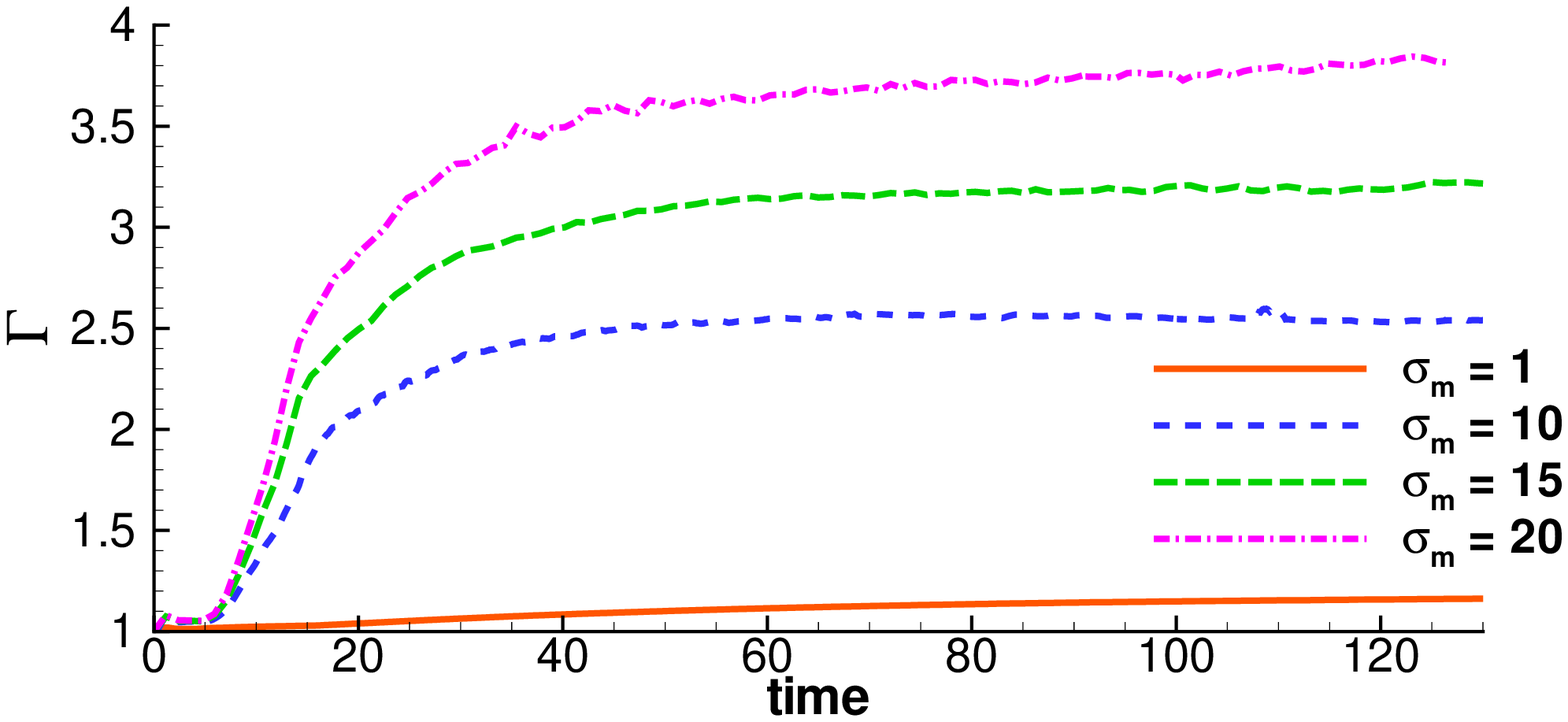,width=8.0cm}
\psfig{file=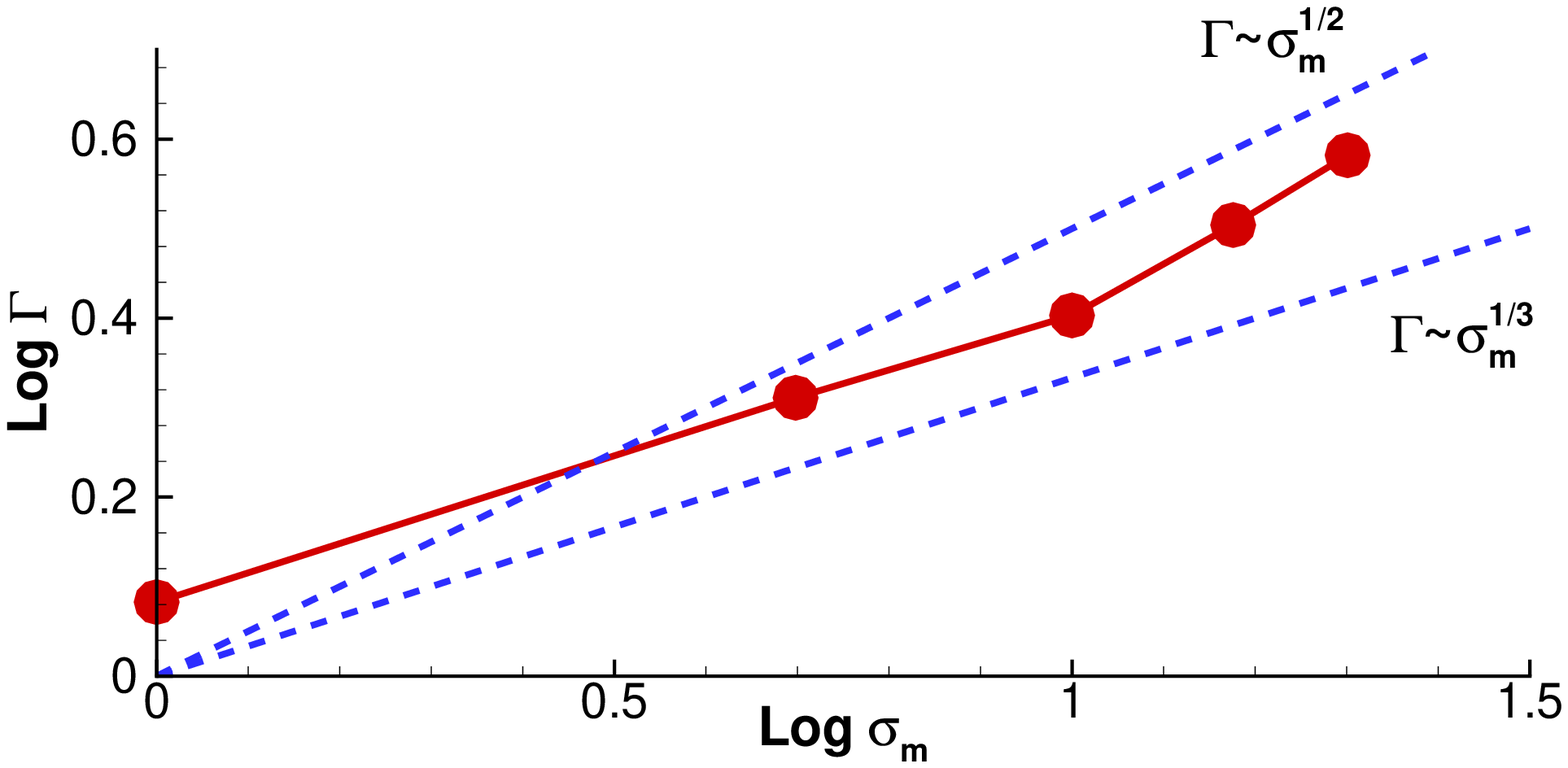,width=8.0cm}
\caption{Time evolution of the magnetic energy (top panel)
  and of the Lorentz factor (middle panel) for
  the first four models reported in Tab.~\ref{tab1}
  with increasing magnetizations $\sigma_m$.
  The bottom panel shows the dependence of the Lorentz
  factor on the magnetization in the plane $\log\sigma_m-\log\Gamma$.}
\label{Lorentz_factor_versus_time_different_beta}
\end{figure}
\begin{figure}
\centering
\hspace{0.2cm}
{\includegraphics[angle=0,width=4.0cm,height=10.0cm]{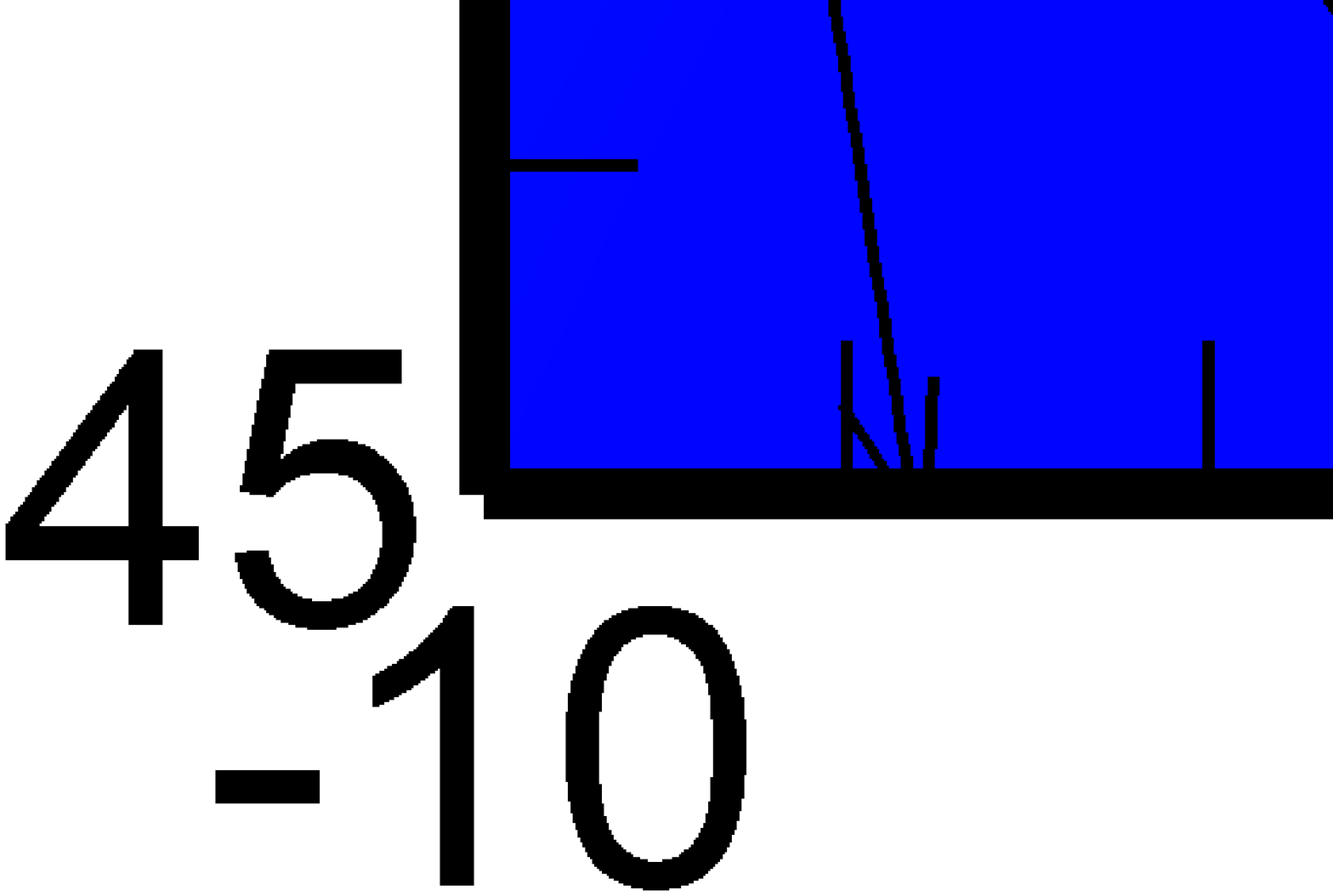}}
{\includegraphics[angle=0,width=4.0cm,height=10.0cm]{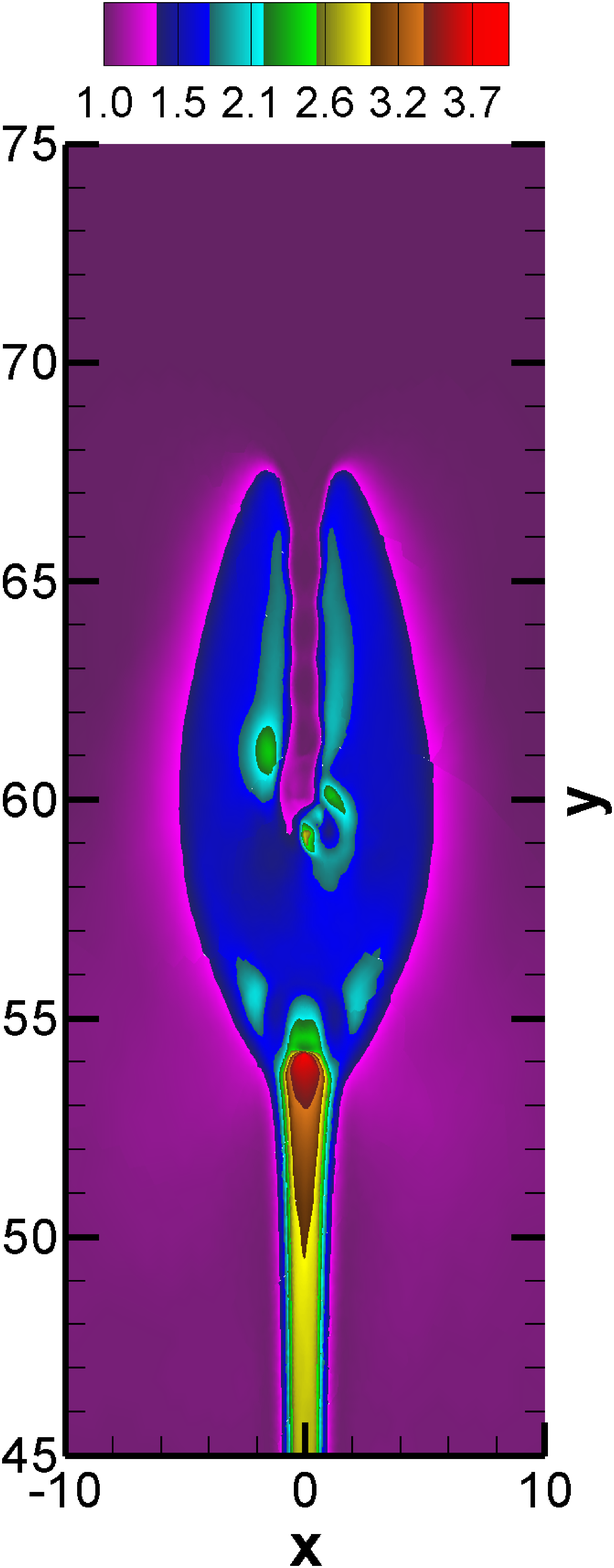}}
\caption{Color map of the rest mass density (left panel) and
  of the Lorentz factor (right panel) 
  for the  model \texttt{2D-m20-S2.82e5}
  with $\sigma_m=20$, $S=2.82\times10^{5}$ at time $t=90$.}
\label{displacement_map}
\end{figure}
In a first series of simulations we have considered the case of
an isotropic Ohm law, namely adopting Eq.~({\ref{ohmlaw}) with
$\xi=0$, concentrating on the effects that an increasing
magnetization produces on the system. 
Fig.~\ref{rho_plus_field_map} shows the rest mass
density and the magnetic field lines for the model 
\texttt{2D-m1-S1.68e5} with $\sigma_m=1.0$ (left panel) and
for the model \texttt{2D-m20-S2.82e5} with $\sigma_m=20$
(right panel). The two panels, which show snapshots of
the two models at the same time $t=90$, confirm the essential
features of a Petschek-type
relativistic reconnection, as also reported  by 
other
authors~\citep{Watanabe2006,Zenitani2009,Zenitani2009_guide}.
Namely, magnetic reconnection is triggered in the
region of anomalous resistivity, producing the typical
$X-$type topology of the magnetic field. As a result, 
magnetic energy is
converted into both thermal and kinetic energy, and 
two plasmoids moving in opposite directions,  
and corresponding to the two high-density
regions of the figure, are accelerated along the direction of the
magnetic field. The magnetic tension, in fact,
is responsible for the collimation of the flow. 
The plasmoid is highly compressed 
by plasma with much higher velocity
and lower density (see the discussion about
Fig.~\ref{displacement_map} below) 
and its rest mass density increases with
increasing magnetization.
A complementary information to that of
Fig.~\ref{rho_plus_field_map} is provided by 
Fig.~\ref{Lorentz_factor_versus_time_different_beta}. The
top and the middle panels show, respectively,
the time evolution of the magnetic energy
(normalized to its initial value)
and of the Lorentz factor of the plasmoid, which is also
the maximum Lorentz factor
monitored over the grid.
The dissipated magnetic energy produces an increase of
both the thermal and kinetic energy. The
latter results in the acceleration of the plasmoid,
which is more efficient for
higher magnetizations. However, as the region around the
anomalous resistivity becomes more and more rarefied, the
conversion of magnetic energy into thermal energy becomes
more efficient than the conversion into kinetic energy
and the Lorentz factor reaches a saturation.
The small wiggles visible in the Lorentz factor curves,
particularly at high magnetizations, are due to impulsive
reconnection episodes, taking place in very small and
localized regions around the plasmoid.
The bottom panel of
Fig.~\ref{Lorentz_factor_versus_time_different_beta}, on
the other hand, shows the maximum Lorentz factor reached
as a function of the magnetization parameter. 
Our results show that, while  at low magnetizations the Lorentz
factor obeys a dependence of the kind $\Gamma\propto
\sigma^{1/3}$, at higher magnetizations the dependence is
close to the theoretical prediction given by 
$\Gamma\propto \sigma_m^{1/2}$~\citep{Lyubarsky2005,Zenitani2009}.

The reconnection rate, namely the speed at which the
reconnection process takes place and that 
we have computed as\footnote{See
\citet{Zenitani2009} for alternative definitions of the 
reconnection rate.} $r=E_z/B_0$, is also
strongly dependent on the magnetization, as firstly
noticed by \citet{Watanabe2006}.
We have
followed its temporal evolution and found that it
reaches a maximum around $t\sim 20-30$, while decreasing
asymptotically after that to reach a stationary value.
The maximum reconnection rate
is $r\sim0.07$ and $r\sim 0.2$ for models with
$\sigma_m=1.0$ and $\sigma_m=20$, respectively.
It should be noted that the
regions of maximum rest mass density and of maximum
Lorentz factor do not  match exactly. The mismatch is
reported in Fig.~\ref{displacement_map}, which provides a
focus of the upper plasmoid that is visible in the right panel of 
Fig.~\ref{rho_plus_field_map}. The left and the right panels of
Fig.~\ref{displacement_map} show the rest mass density
and the Lorentz factor, respectively. As it is apparent
from this figure, the plasma reaches its maximum velocity
at the basis of the plasmoid in a very rarefied region. 
On the other
hand, the portion of the plasmoid with maximum rest
mass density has very low Lorentz
factor. As a result, a tiny bow shock is 
produced at the basis of the plasmoid, confirming
similar findings by \citet{Zenitani2010} who performed a
detailed analysis of the
generation of slow shocks around the plasmoids.

In a second series of simulations, we have analyzed the
dependence of the reconnection process on the background Lundquist
number $S$, while keeping the same
peak value of the anomalous resistivity $\eta_{i0}=1.0$
and of the magnetization parameter $\sigma_m=10$.
It is worth recalling that high values of $S$,
corresponding to higher background conductivities,
represent a challenge for the numerical
scheme, as the stiffness  of the resistive
magnetohydrodynamics equations becomes more severe. 
However, a higher
resistivity jump between the background and the anomalous
resistivity reproduces physical conditions where the plasma
conductivity changes sharply over small distance scales,
as expected in realistic conditions.
The top panel of Fig.~\ref{Lorentz_factor_versus_time_different_Lundquist}
shows that, when increasing the Lundquist number from $S=2.68\times10^{5}$ to
$S=2.68\times10^{8}$, 
the asymptotic Lorentz factor reached by the
plasmoid is smaller. 
The explanation of this effect is
once again to be found in the efficiency of magnetic
energy conversion. When there is a higher  conductivity
jump between the central hot spot and the background, in fact,
the fraction of magnetic energy that converts 
into thermal energy increases. 
This is shown in the middle panel of
Fig.~\ref{Lorentz_factor_versus_time_different_Lundquist},
which reports the time evolution of the specific internal
energy of the plasma, which is higher for higher
Lundquist numbers. The bottom panel of
Fig.~\ref{Lorentz_factor_versus_time_different_Lundquist},
on the other hand, shows the evolution of the
reconnection rate $r=E_z/B_0$, whose peak is a mild
decreasing function of the Lundquist number. 
As it is evident from the top panel of
Fig.~\ref{Lorentz_factor_versus_time_different_Lundquist},
the Lorentz factor for the model with
$S=2.68\times10^{8}$ manifests an irregular evolution
after $t=70$, which is the signature of the development of a
magnetic instability.

\begin{figure}
\psfig{file=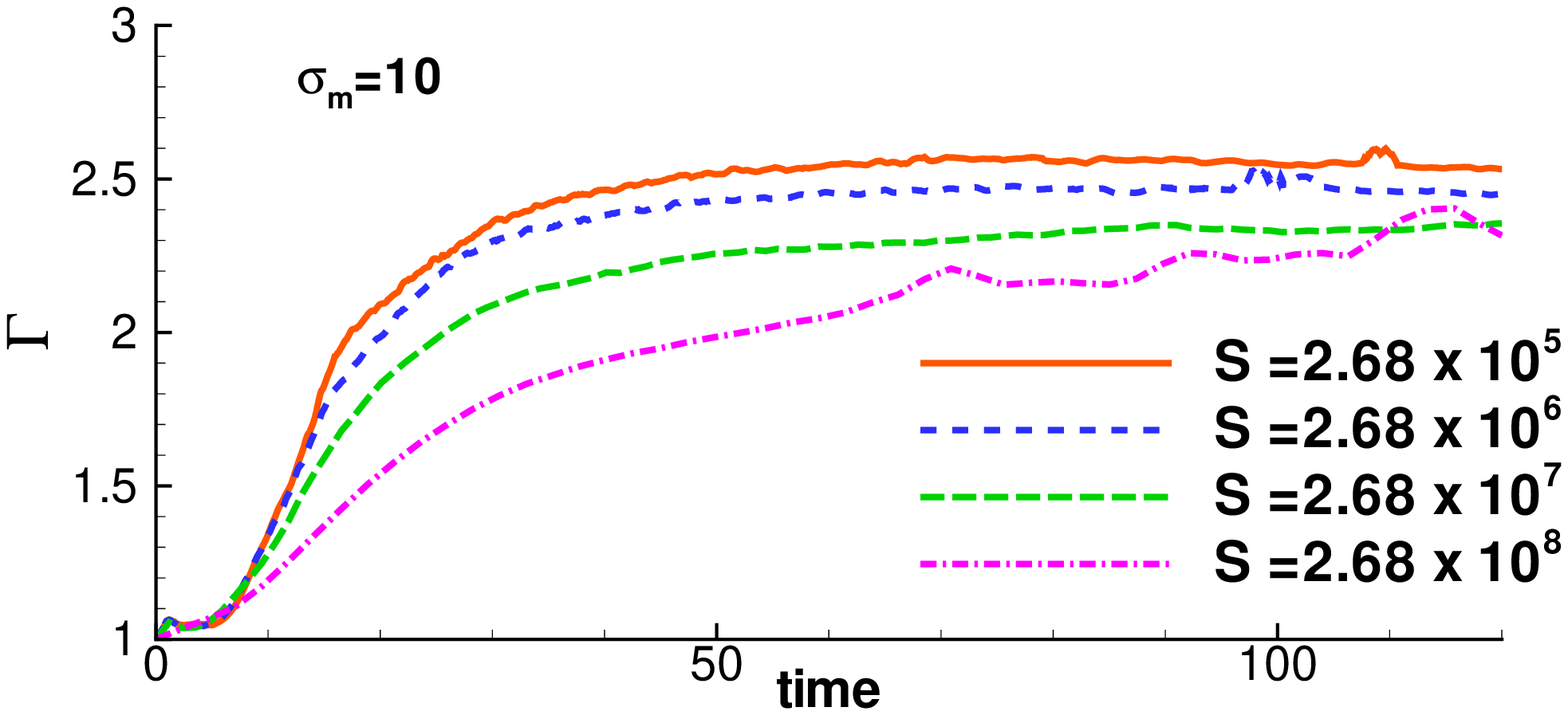,width=8.0cm}
\psfig{file=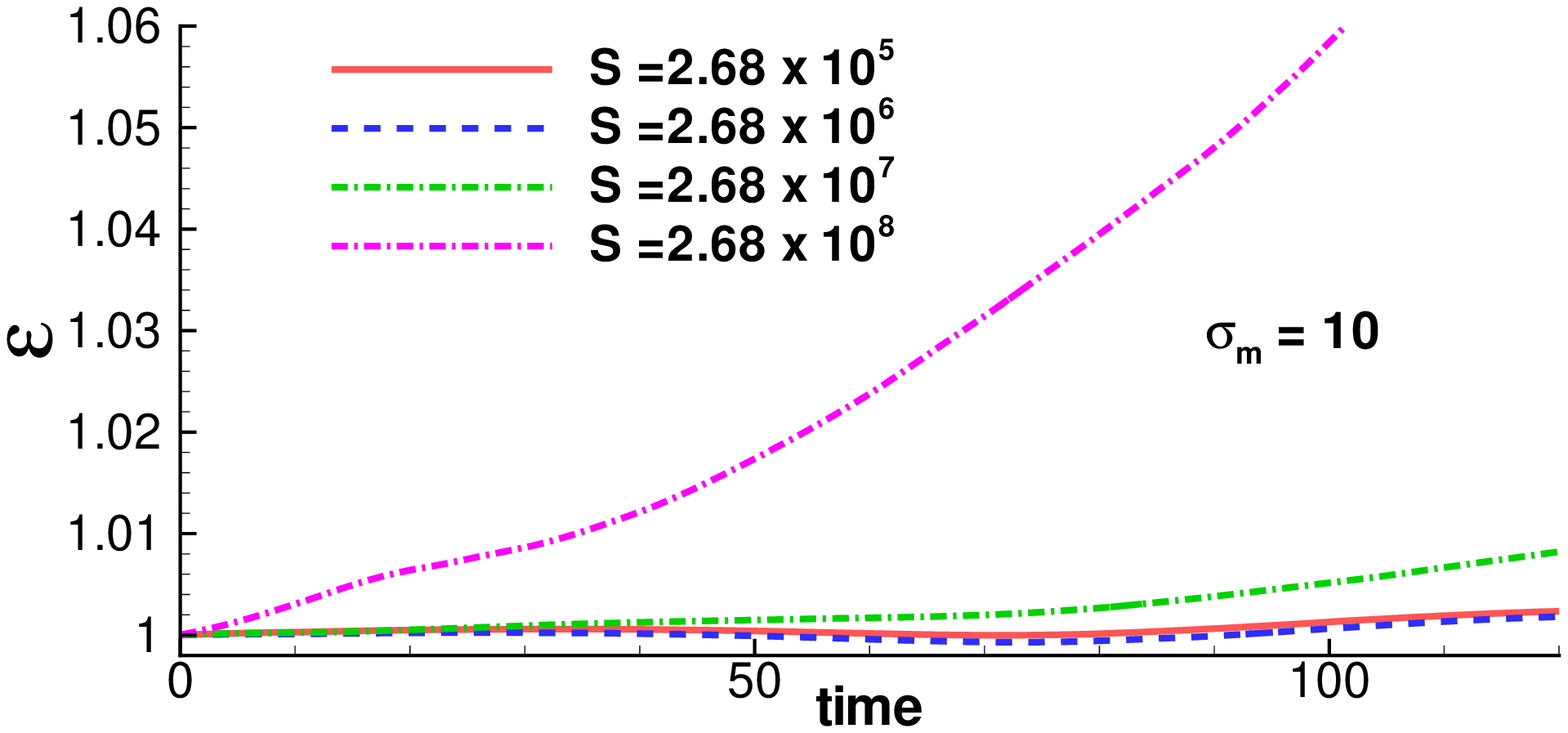,width=8.0cm}
\psfig{file=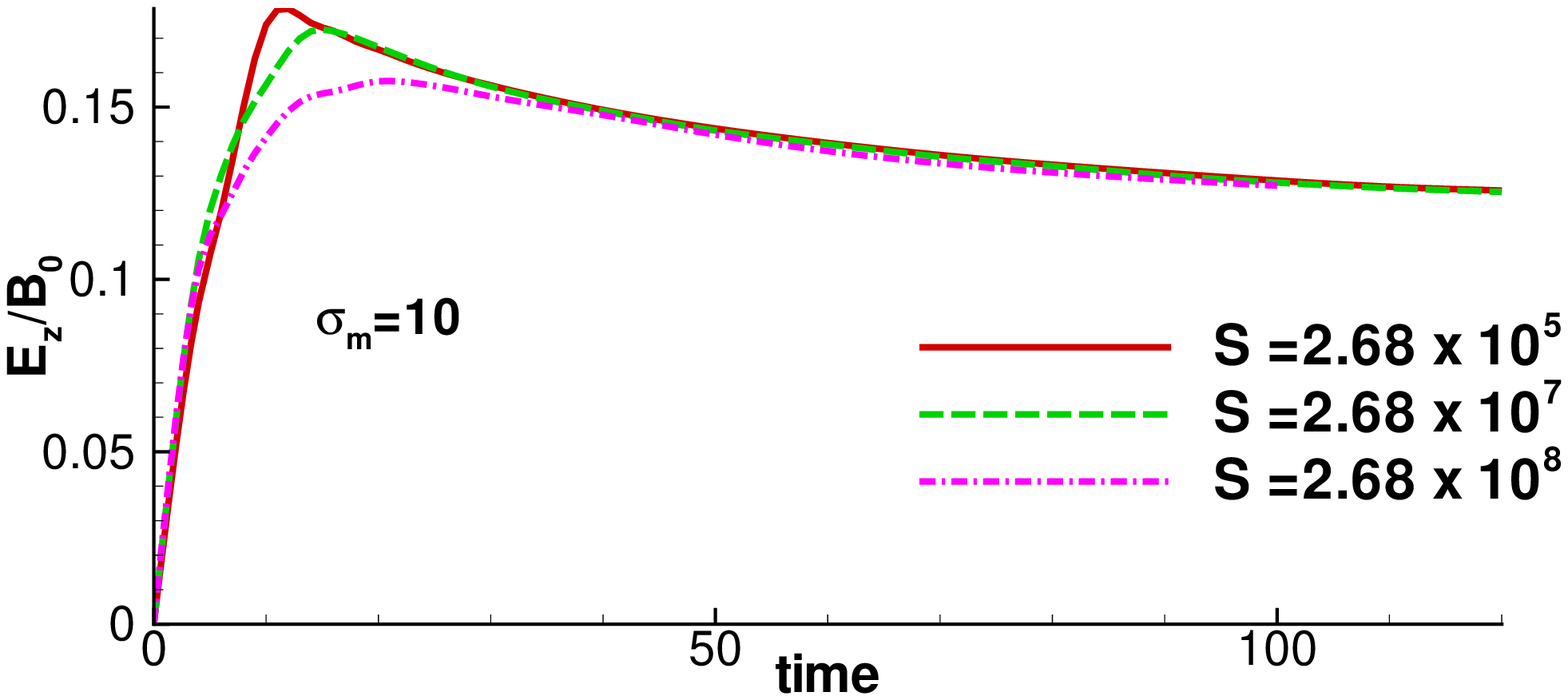,width=8.0cm}
\caption{Time evolution of the Lorentz factor (top panel),
  of the thermal energy (middle panel) and of the
  reconnection rate $E_z/B_0$ (bottom panel) for 
  models having the same
  magnetization $\sigma_m=10.0$ but different Lundquist
  numbers.
  }
\label{Lorentz_factor_versus_time_different_Lundquist}
\end{figure}
\begin{figure}
\centering
\hspace{0.2cm}
{\includegraphics[angle=0,width=4.0cm,height=10.0cm]{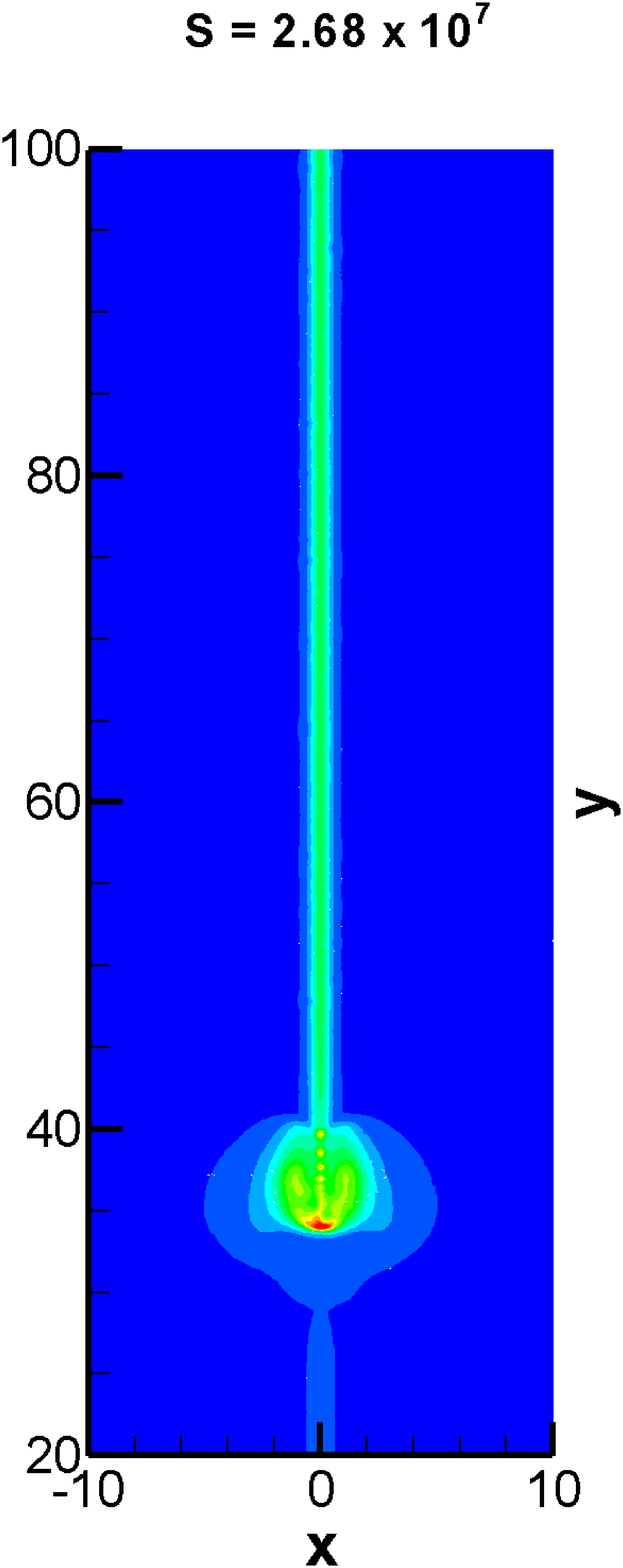}}
{\includegraphics[angle=0,width=4.0cm,height=10.0cm]{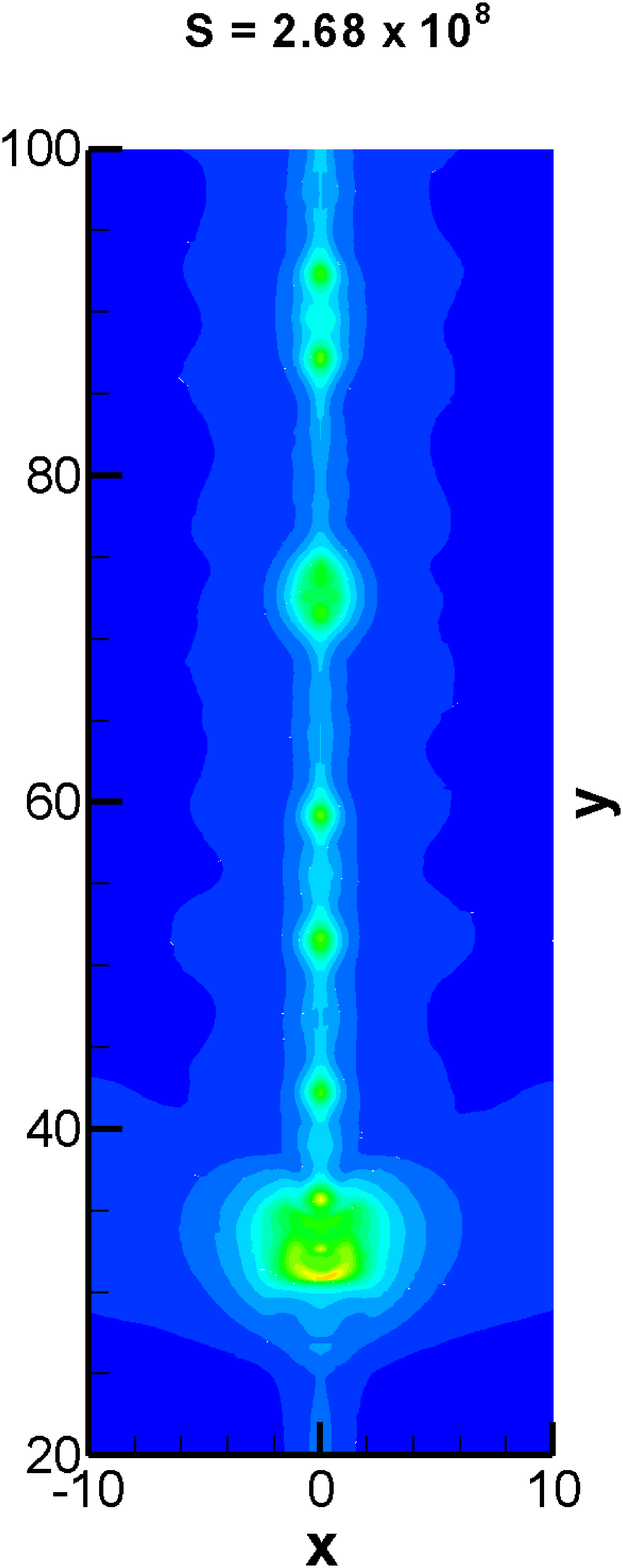}}
\caption{Generation of plasmoid chains in very high
  Lundquist numbers configurations. Left and right
  panels report, respectively,
  the color map of the rest mass density  at time
  $t=60$ for two models having $S=2.68\times 10^7$
  and $S=2.68\times 10^8$.}
\label{plsmoid_chain}
\end{figure}

This effect is reported in
Fig.~\ref{plsmoid_chain}, showing the rest mass density
in two models having different Lundquist
numbers. No sign of instability is visible in simulations
with Lundquist numbers as large as $S\sim10^7$ (left panel), while
a chain of magnetic islands is produced for $S>S_c\sim10^8$ (right panel). This
instability, which resembles a tearing instability,
was investigated through a linear analysis by
\citet{Loureiro2007}, and subsequently confirmed via
numerical simulations in the Newtonian framework 
by \citet{Samtaney2009}. Our results, 
combined with those by \citet{Samtaney2009}, 
indicate that, in
the transition to the relativistic regime, the critical
Lundquist number increases from $S_c\gtrsim 10^4$ to
$S_c\gtrsim 10^8$. Such a conclusion may have deep
implications for high Lundquist number reconnection in
realistic astrophysical conditions, and it will 
deserve further investigations.

Finally, in a third series of simulations we have considered
the effects of the full anisotropic
Ohm law given by Eq.~\eqref{ohmlaw} with $\xi\neq0$.
In this case, a component of the magnetic field
perpendicular to the $x-y$ plane must be introduced,
otherwise the term $\vec E \cdot \vec B$ in
(\ref{ohmlaw}) remains zero. Hence, we have investigated
anisotropic effects in three spatial dimensions and
in configurations with $B_z\neq
0$, the so called ``guide field''
configurations~\citep{Lyubarsky2005}, 
for
which we have chosen the value  $B_z=0.5 B_0$. 
As a representative example, Fig.~\ref{3D_figure} shows
the magnetic field $B^2$ on the slices $x=0$ and $z=0$
for the model \texttt{3D-m1.25-Bz0.5} at time $t=90$.
The diffusion of the magnetic field is clearly visible in
the region around the anomalous resistivity.
It has to be remarked that, within our single fluid
approximation, inertial resistivity effects like those
encountered by \cite{Zenitani2009_guide} cannot be captured
and are not discussed here.
Simulations with the guide field
configuration in combination with the anisotropic Ohm law
turned out to be very challenging for the numerical
scheme. In particular, strong magnetizations could not be
reached and the results that we show here are limited to
the case $\sigma_m\sim 1$. 
When discussing the effects introduced by 
an anisotropic Ohm law, we first need to distinguish
them from those produced by the guide-field.  
By using a two-fluid approach, \cite{Zenitani2009_guide}
concluded that the guide field modifies the composition of
the output energy flux, which, as the guide field
increases, 
changes from being enthalpy
dominated to be Poynting dominated. Even within a
single fluid approach, we confirm here a similar
finding. This is shown in
Fig.~\ref{lorentz_factor_isotropic_versus_anisotropic},
where we compare the behavior of three different configurations
each of which with an effective $\sigma_m=1.25$.
The solid red and the dashed blue line,
in fact, refer, respectively, 
to a 3D simulation without the guide magnetic field 
and to a 3D simulation
with the guide magnetic field, but with an otherwise isotropic
Ohm law. In the case of a non zero guide field, the
Lorentz factor is substantially smaller, while the decay
of the magnetic energy, not reported here, is
correspondingly slower. This is consistent with what
reported by \cite{Zenitani2009_guide}, who found that the
introduction of the guide field has the net effect of
diminishing the bilk kinetic energy.
Having clarified this, we have increased the
parameter $\xi$ to establish the extent to which results
are affected by anisotropic Ohm law (green long-dashed line).
The present implementation of the numerical scheme
does not allow to treat 
the anisotropic parameter $\xi$ as a true free parameter,
and we have therefore concentrated on
a mildly anisotropic regime. 
When $\xi^2$ is increased from $0$ (isotropic case) to $0.5$,
the Lorentz factor indeed grows, but
by $1\%$ only after $t=100$. Intuitively, this effect can
be explained in terms of the increased Lorentz force on
the plasma, because of the extra current term on the
right hand side of \eqref{ohmlaw2}. More work is needed
to analyze the anisotropic
regime under more realistic conditions.

\begin{figure}
\psfig{file=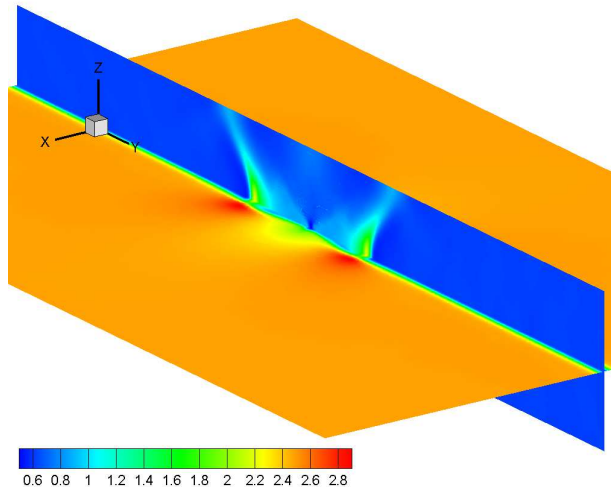,width=8.0cm}
\caption{Magnetic field $B^2$ on the slice  $x=0$ and on
  the slice $z=0$ for the model
  \texttt{3D-m1.25-Bz0.5} at time $t=90$.}
\label{3D_figure}
\end{figure}
\begin{figure}
\psfig{file=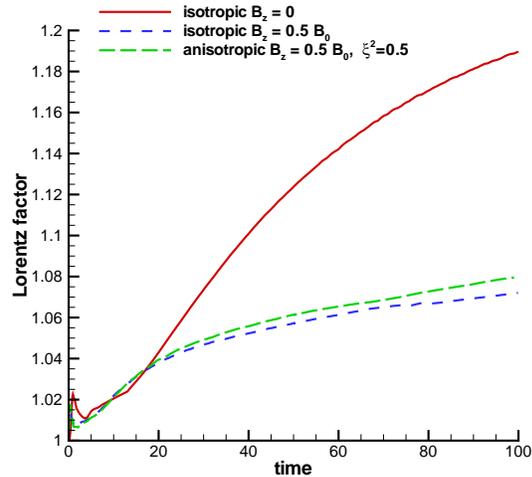,width=8.0cm}
\caption{Combined effects of the guide field and of
  anisotropic Ohm law on the
  time evolution of the Lorentz factor. All of the models
  shown have $\sigma_m=1.25$. See text for explanation.}
\label{lorentz_factor_isotropic_versus_anisotropic}
\end{figure}
%
%

\section{Conclusions}
\label{sec:Conclusions}

We have investigated
the dynamics of High Lundquist number relativistic magnetic
reconnection, 
by performing two dimensional and three dimensional
magnetohydrodynamics simulations 
of a Petschek type configuration.
By resorting to high order
discontinuous Galerkin methods as proposed by~\citet{Dumbser2009},
we have
found that Lorentz factors up to $\sim4$
can be obtained for plasma parameters $\sigma_m$ up to
$20$ in systems with Lundquist number $S\sim10^5$.
However, such values of the Lorentz factor decrease when
$S$ is increased as a result of a reduced 
background resistivity. When $S$ is larger than a
critical value $S_c\sim 10^8$, 
the Sweet-Parker layer becomes
unstable, generating a chain of secondary magnetic
islands. This effect deserves additional theoretical
investigations, 
in view of the fact that
a similar instability 
has already been reported in the
Newtonian framework~\citep{Samtaney2009}, but at much lower critical
Lundquist numbers, i.e. for $S>S_c\sim 10^4$.

We have also shown that, when an
anisotropic Ohm law is adopted, both the reconnection rate
and  the Lorentz factor of the
accelerated plasmoid are slightly increased.
Although this increase is small
and within a few percent with respect to the isotropic
Ohm law, strongly anisotropic regimes remain challenging
even for our advanced numerical scheme.
We plan to improve
on these limitations in our  
future applications of relativistic magnetic reconnection
to the curved spacetime around a neutron star,
where an anisotropic Ohm law can play a
substantial role.

Finally, it is worth mentioning that 
pure magnetohydrodynamics reconnection occurs in
collisional plasmas, while the transition to
collisionless reconnection is likely to enhance the reconnection
rate~\citep{Uzdensky2007,Uzdensky2010}, and, presumably, also the
acceleration properties. 
This possibility, which is related to some recent results
of plasma physics \citep{Hesse2009,Bessho2007},
 has been explored by
\citet{Goodman2008} in accretion disc coronae and by
\citet{McKinney2010} to trigger dissipation in Gamma-Ray
Burst jets. 
The numerical exploration of
collisionless reconnection has received less
attention\footnote{Also note that \citet{Lazarian2010}, 
after solving numerically a reduced set of equations, showed that
reconnection in a turbulent fluid occurs at a rate
comparable to the rms velocity of the turbulence,
irrespective of the degree of collisionality.}  
and it will represent another direction of our future analysis.

\section*{Acknowledgments}
O.Z. is grateful to Luciano Rezzolla and to Serguei
Komissarov for very useful discussions. We would like to
thank an anonymous referee for very useful comments.
Numerical simulations where performed on the 
National Supercomputer HLRB-II based on  SGI's Altix 4700
platform and installed at Leibniz-Rechenzentrum.
Partial support comes from the DFG grant SFB/Transregio
7, by ``CompStar'', a Research Net-working Programme of the
European Science Foundation. 

\bibliographystyle{mn2e}
\bibliography{aeireferences}

\begin{thebibliography}{}

\bibitem[\protect\citeauthoryear{{Barkov} \& {Komissarov}}{{Barkov} \&
  {Komissarov}}{2010}]{Barkov2010}
{Barkov} M.~V.,  {Komissarov} S.~S.,  2010, Mon. Not. R. Astron. Soc., 401,
  1644

\bibitem[\protect\citeauthoryear{{Bekenstein} \& {Oron}}{{Bekenstein} \&
  {Oron}}{1978}]{Bekenstein1978}
{Bekenstein} J.~D.,  {Oron} E.,  1978, Phys. Rev. D, 18, 1809

\bibitem[\protect\citeauthoryear{{Bessho} \& {Bhattacharjee}}{{Bessho} \&
  {Bhattacharjee}}{2007}]{Bessho2007}
{Bessho} N.,  {Bhattacharjee} A.,  2007, Physics of Plasmas, 14, 056503

\bibitem[\protect\citeauthoryear{{Blackman} \& {Field}}{{Blackman} \&
  {Field}}{1994}]{Blackman1994}
{Blackman} E.~G.,  {Field} G.~B.,  1994, Phys. Rev. Lett., 72, 494

\bibitem[\protect\citeauthoryear{{Dedner}, {Kemm}, {Kr{\"o}ner}, {Munz},
  {Schnitzer} \& {Wesenberg}}{{Dedner} et~al.}{2002}]{Dedner2002}
{Dedner} A.,  {Kemm} F.,  {Kr{\"o}ner} D.,  {Munz} C.-D.,  {Schnitzer} T.,
  {Wesenberg} M.,  2002, J. Comput. Phys., 175, 645

\bibitem[\protect\citeauthoryear{{Del Zanna}, {Zanotti}, {Bucciantini} \&
  {Londrillo}}{{Del Zanna} et~al.}{2007}]{DelZanna2007}
{Del Zanna} L.,  {Zanotti} O.,  {Bucciantini} N.,    {Londrillo} P.,  2007,
  Astron. Astrophys., 473, 11

\bibitem[\protect\citeauthoryear{{di Matteo}}{{di Matteo}}{1998}]{diMatteo1998}
{di Matteo} T.,  1998, Mon. Not. R. Astron. Soc., 299, L15

\bibitem[\protect\citeauthoryear{{Drenkhahn} \& {Spruit}}{{Drenkhahn} \&
  {Spruit}}{2002}]{Drenkhahn2002}
{Drenkhahn} G.,  {Spruit} H.~C.,  2002, Astron. Astrophys., 391, 1141

\bibitem[\protect\citeauthoryear{{Dumbser}, {Balsara}, {Toro} \&
  {Munz}}{{Dumbser} et~al.}{2008}]{DBTM2008}
{Dumbser} M.,  {Balsara} D.~S.,  {Toro} E.~F.,    {Munz} C.-D.,  2008, Journal
  of Computational Physics, 227, 8209

\bibitem[\protect\citeauthoryear{{Dumbser}, {Enaux} \& {Toro}}{{Dumbser}
  et~al.}{2008}]{DET2008}
{Dumbser} M.,  {Enaux} C.,    {Toro} E.~F.,  2008, Journal of Computational
  Physics, 227, 3971

\bibitem[\protect\citeauthoryear{{Dumbser} \& {Zanotti}}{{Dumbser} \&
  {Zanotti}}{2009}]{Dumbser2009}
{Dumbser} M.,  {Zanotti} O.,  2009, Journal of Computational Physics, 228, 6991

\bibitem[\protect\citeauthoryear{{Goodman} \& {Uzdensky}}{{Goodman} \&
  {Uzdensky}}{2008}]{Goodman2008}
{Goodman} J.,  {Uzdensky} D.,  2008, Astrophys. J., 688, 555

\bibitem[\protect\citeauthoryear{{Gruzinov}}{{Gruzinov}}{2005}]{Gruzinov2005}
{Gruzinov} A.,  2005, Phys. Rev. Lett., 94, 021101

\bibitem[\protect\citeauthoryear{Harten, Engquist, Osher \&
  Chakravarthy}{Harten et~al.}{1987}]{eno}
Harten A.,  Engquist B.,  Osher S.,    Chakravarthy S.,  1987, Journal of
  Computational Physics, 71, 231

\bibitem[\protect\citeauthoryear{{Hesse}, {Zenitani}, {Kuznetsova} \&
  {Klimas}}{{Hesse} et~al.}{2009}]{Hesse2009}
{Hesse} M.,  {Zenitani} S.,  {Kuznetsova} M.,    {Klimas} A.,  2009, Physics of
  Plasmas, 16, 102106

\bibitem[\protect\citeauthoryear{{Jaroschek}, {Lesch} \&
  {Treumann}}{{Jaroschek} et~al.}{2004}]{Jaroschek2004}
{Jaroschek} C.~H.,  {Lesch} H.,    {Treumann} R.~A.,  2004, Astrophys. J. Lett,
  605, L9

\bibitem[\protect\citeauthoryear{{Kandus} \& {Tsagas}}{{Kandus} \&
  {Tsagas}}{2008}]{Kandus2008}
{Kandus} A.,  {Tsagas} C.~G.,  2008, Mon. Not. R. Astron. Soc., 385, 883

\bibitem[\protect\citeauthoryear{{Kirk} \& {Skj{\ae}raasen}}{{Kirk} \&
  {Skj{\ae}raasen}}{2003}]{Kirk2003}
{Kirk} J.~G.,  {Skj{\ae}raasen} O.,  2003, Astrophys. J., 591, 366

\bibitem[\protect\citeauthoryear{Klaij, der Vegt \& der Ven}{Klaij
  et~al.}{2006}]{KlaijVanDerVegt}
Klaij C.,  der Vegt J.~V.,    der Ven H.~V.,  2006, Journal of Computational
  Physics, 217, 589

\bibitem[\protect\citeauthoryear{{Komissarov}}{{Komissarov}}{1997}]{Komissarov%
1997}
{Komissarov} S.~S.,  1997, Physics Letters A, 232, 435

\bibitem[\protect\citeauthoryear{{Komissarov}}{{Komissarov}}{2007}]{Komissarov%
2007}
{Komissarov} S.~S.,  2007, Mon. Not. R. Astron. Soc., 382, 995

\bibitem[\protect\citeauthoryear{{Lazarian}, {Kowal}, {Vishniac} \& {de Gouveia
  dal Pino}}{{Lazarian} et~al.}{2010}]{Lazarian2010}
{Lazarian} A.,  {Kowal} G.,  {Vishniac} E.,    {de Gouveia dal Pino} E.,  2010,
  ArXiv e-prints

\bibitem[\protect\citeauthoryear{{Loureiro}, {Schekochihin} \&
  {Cowley}}{{Loureiro} et~al.}{2007}]{Loureiro2007}
{Loureiro} N.~F.,  {Schekochihin} A.~A.,    {Cowley} S.~C.,  2007, Physics of
  Plasmas, 14, 100703

\bibitem[\protect\citeauthoryear{{Lyubarsky}}{{Lyubarsky}}{2005}]{Lyubarsky200%
5}
{Lyubarsky} Y.~E.,  2005, Mon. Not. Roy. Astr. Soc., 358, 113

\bibitem[\protect\citeauthoryear{{Lyutikov}}{{Lyutikov}}{2003}]{Lyutikov2003}
{Lyutikov} M.,  2003, Mon. Not. R. Astron. Soc., 346, 540

\bibitem[\protect\citeauthoryear{{Lyutikov}}{{Lyutikov}}{2006}]{Lyutikov2006}
{Lyutikov} M.,  2006, Mon. Not. R. Astron. Soc., 367, 1594

\bibitem[\protect\citeauthoryear{{Lyutikov} \& {Uzdensky}}{{Lyutikov} \&
  {Uzdensky}}{2003}]{Lyutikov2003b}
{Lyutikov} M.,  {Uzdensky} D.,  2003, Astrophysical Journal, 589, 893

\bibitem[\protect\citeauthoryear{{McKinney} \& {Uzdensky}}{{McKinney} \&
  {Uzdensky}}{2010}]{McKinney2010}
{McKinney} J.~C.,  {Uzdensky} D.~A.,  2010, ArXiv e-prints

\bibitem[\protect\citeauthoryear{{Nalewajko}, {Giannios}, {Begelman},
  {Uzdensky} \& {Sikora}}{{Nalewajko} et~al.}{2011}]{Nalewajko2011}
{Nalewajko} K.,  {Giannios} D.,  {Begelman} M.~C.,  {Uzdensky} D.~A.,
  {Sikora} M.,  2011, Mon. Not. R. Astron. Soc., p.~188

\bibitem[\protect\citeauthoryear{{Palenzuela}, {Lehner}, {Reula} \&
  {Rezzolla}}{{Palenzuela} et~al.}{2009}]{Palenzuela:2008sf}
{Palenzuela} C.,  {Lehner} L.,  {Reula} O.,    {Rezzolla} L.,  2009, Mon. Not.
  R. Astron. Soc., 394, 1727

\bibitem[\protect\citeauthoryear{{P{\'e}tri} \& {Lyubarsky}}{{P{\'e}tri} \&
  {Lyubarsky}}{2007}]{Petri2007}
{P{\'e}tri} J.,  {Lyubarsky} Y.,  2007, Astron. Astrophys., 473, 683

\bibitem[\protect\citeauthoryear{{Radice} \& {Rezzolla}}{{Radice} \&
  {Rezzolla}}{2011}]{Radice2011}
{Radice} D.,  {Rezzolla} L.,  2011, ArXiv e-prints, 1103.2426

\bibitem[\protect\citeauthoryear{{Rezzolla}, {Giacomazzo}, {Baiotti}, {Granot},
  {Kouveliotou} \& {Aloy}}{{Rezzolla} et~al.}{2011}]{Rezzolla:2011}
{Rezzolla} L.,  {Giacomazzo} B.,  {Baiotti} L.,  {Granot} J.,  {Kouveliotou}
  C.,    {Aloy} M.~A.,  2011, Astrophys. Journ. Lett., 732, L6

\bibitem[\protect\citeauthoryear{{Samtaney}, {Loureiro}, {Uzdensky},
  {Schekochihin} \& {Cowley}}{{Samtaney} et~al.}{2009}]{Samtaney2009}
{Samtaney} R.,  {Loureiro} N.~F.,  {Uzdensky} D.~A.,  {Schekochihin} A.~A.,
  {Cowley} S.~C.,  2009, Physical Review Letters, 103, 105004

\bibitem[\protect\citeauthoryear{{Schopper}, {Lesch} \& {Birk}}{{Schopper}
  et~al.}{1998}]{Schopper1998}
{Schopper} R.,  {Lesch} H.,    {Birk} G.~T.,  1998, Astron. Astrophys., 335, 26

\bibitem[\protect\citeauthoryear{{Tanuma}, {Yokoyama}, {Kudoh} \&
  {Shibata}}{{Tanuma} et~al.}{2003}]{Tanuma2003}
{Tanuma} S.,  {Yokoyama} T.,  {Kudoh} T.,    {Shibata} K.,  2003, Astrophys.
  J., 582, 215

\bibitem[\protect\citeauthoryear{{Tolstykh}, {Semenov}, {Biernat}, {Heyn} \&
  {Penz}}{{Tolstykh} et~al.}{2007}]{Tolstykh2007}
{Tolstykh} Y.~V.,  {Semenov} V.~S.,  {Biernat} H.~K.,  {Heyn} M.~F.,    {Penz}
  T.,  2007, Advances in Space Research, 40, 1538

\bibitem[\protect\citeauthoryear{{Uzdensky}}{{Uzdensky}}{2003}]{Uzdensky2003}
{Uzdensky} D.~A.,  2003, Astrophysical Journal, 598, 446

\bibitem[\protect\citeauthoryear{{Uzdensky}}{{Uzdensky}}{2007}]{Uzdensky2007}
{Uzdensky} D.~A.,  2007, Astrop. J., 671, 2139

\bibitem[\protect\citeauthoryear{{Uzdensky}, {Loureiro} \&
  {Schekochihin}}{{Uzdensky} et~al.}{2010}]{Uzdensky2010}
{Uzdensky} D.~A.,  {Loureiro} N.~F.,    {Schekochihin} A.~A.,  2010, Physical
  Review Letters, 105, 235002

\bibitem[\protect\citeauthoryear{van~der Vegt \& van~der Ven}{van~der Vegt \&
  van~der Ven}{2002}]{spacetimedg1}
van~der Vegt J. J.~W.,  van~der Ven H.,  2002, Journal of Computational
  Physics, 182, 546

\bibitem[\protect\citeauthoryear{{Watanabe} \& {Yokoyama}}{{Watanabe} \&
  {Yokoyama}}{2006}]{Watanabe2006}
{Watanabe} N.,  {Yokoyama} T.,  2006, Astrophys. J. Lett., 647, L123

\bibitem[\protect\citeauthoryear{{Zenitani}, {Hesse} \& {Klimas}}{{Zenitani}
  et~al.}{2009a}]{Zenitani2009_guide}
{Zenitani} S.,  {Hesse} M.,    {Klimas} A.,  2009a, Astrophysical Journal, 705,
  907

\bibitem[\protect\citeauthoryear{{Zenitani}, {Hesse} \& {Klimas}}{{Zenitani}
  et~al.}{2009b}]{Zenitani2009}
{Zenitani} S.,  {Hesse} M.,    {Klimas} A.,  2009b, Astrophysical Journal, 696,
  1385

\bibitem[\protect\citeauthoryear{{Zenitani}, {Hesse} \& {Klimas}}{{Zenitani}
  et~al.}{2010}]{Zenitani2010}
{Zenitani} S.,  {Hesse} M.,    {Klimas} A.,  2010, Astrophysical Journal Lett.,
  716, L214

\bibitem[\protect\citeauthoryear{Zumbusch}{Zumbusch}{2009}]{zumbusch_2009_fed}
Zumbusch G.,  2009, Classical and Quantum Gravity, 26, 175011

\end{thebibliography}


\bsp

\label{lastpage}

\appendix
\section[]{Derivation of Eq.~(21)}
\label{appendixA}

We first write the fluid four velocity, the magnetic
field in the comoving frame 
and the electric field in the comoving frame as
\bea
u^\mu&=&\Gamma(v^\mu+n^\mu) \\
e^\mu&=&F^{\mu\nu}u_\nu=\Gamma[ n^\mu (\vec v \cdot \vec
  E) +E^\mu+(\vec v\times \vec B)^\mu] \\
b^\mu&=&F^{\ast\mu\nu}u_\nu=\Gamma[ n^\mu (\vec v \cdot \vec
  B) +B^\mu-(\vec v\times \vec E)^\mu] 
\eea 
where we have used the definitions (\ref{emtensor1}) and
(\ref{emtensor2}) for
the electromagnetic tensor and where we have used the
property
$\epsilon^{\mu\nu\lambda}=\epsilon^{\mu\nu\lambda\kappa}n_\kappa$. 
We then refer to the relation 
\be
\sigma^{\mu\nu}=\sigma(g^{\mu\nu}+\xi^2 b^{\mu}b^{\nu})
\ee
and after noticing that $b^\mu e_\mu=\vec E \cdot \vec B$
(it is a relativistic invariant) we rewrite (\ref{ohm1}) as
\bea
I^\mu&=&q_0\Gamma(v^\mu + n^\mu)+ \sigma \Gamma[(\vec v \cdot
  \vec E)n^\mu + E^\mu + (\vec v \times \vec B)^\mu] +
\nonumber \\
\label{ff0}
&& \sigma\xi^2(\vec v \cdot \vec E)\Gamma
[(\vec v \cdot \vec B)n^\mu + B^\mu - (\vec v \times \vec
  E)^\mu] \ .
\eea
At this point we compare (\ref{ff0}) with $I^\mu=\rho_c
n^\mu + J^\mu$ to find
\bea
\label{111}
\rho_c&=&q_0\Gamma + \sigma\Gamma (\vec v \cdot \vec E) +
\sigma\xi^2 \Gamma (\vec E \cdot \vec B)(\vec v \cdot \vec B) \\
\label{112}
J^\mu&=&q_0\Gamma v^\mu + \sigma\Gamma E^\mu
+\sigma\Gamma(\vec v \times \vec B)^\mu + \nonumber \\
&&\sigma\xi^2(\vec
E \cdot \vec B)\Gamma (B^\mu-(\vec v \times \vec E)^\mu)
\ .
\eea
A last replacement of $q_0$ from (\ref{111}) into
(\ref{112}) allows to derive (\ref{ohmlaw}) of the main text.

\end{document}